\documentclass[11pt,reqno,twoside]{amsart}
\usepackage[latin1]{inputenc}
\usepackage{amssymb,amsfonts,amsmath}
\usepackage[english]{babel}
\usepackage{pstricks}
\usepackage{longtable}
\usepackage{lscape}
\usepackage{booktabs}
\usepackage{caption}
\usepackage{graphicx, subfigure}
\usepackage{natbib}
\usepackage[foot]{amsaddr}
\usepackage{array}
\usepackage{hyperref}
\usepackage{color}
\usepackage{enumerate}

\usepackage{fixltx2e}

\newcommand{\beginsupplement}{%
        \setcounter{table}{0}
        \renewcommand{\thetable}{S\arabic{table}}%
        \setcounter{figure}{0}
        \renewcommand{\thefigure}{S\arabic{figure}}%
     }

\begin{document}

\title[A Mathematical Model For Cell Migration Assays Using A Real-Time Cell Analysis]{A Macroscopic Mathematical Model For Cell Migration Assays Using A Real-Time Cell Analysis}

\author[E. Di Costanzo]{Ezio Di Costanzo$^{1*}$}
\author[V. Ingangi]{Vincenzo Ingangi$^{2,3*}$}
\author[C. Angelini]{Claudia Angelini$^{1}$}
\author[M. F. Carfora]{Maria Francesca Carfora$^{1}$}
\author[M. V. Carriero]{Maria Vincenza Carriero$^{2}$}
\author[R. Natalini]{Roberto Natalini$^{1}$}
\address{$^{1}$Istituto per le Applicazioni del Calcolo ``M. Picone'', Consiglio Nazionale delle Ricerche, Italy.}
\address{$^{2}$ Neoplastic Progression Unit, Department of Experimental Oncology, IRCCS Istituto Nazionale Tumori ``Fondazione G. Pascale'', Naples, Italy.}
\address{$^{3}$ SUN Second University of Naples, Italy.}
\address{$^{*}$ These authors contributed equally to this project and should be considered co-first authors.}
\email{$$e.dicostanzo@na.iac.cnr.it (corresponding author)}
\email{$$vincenzo.ingangi@unina2.it}
\email{$$c.angelini@na.iac.cnr.it}
\email{$$f.carfora@na.iac.cnr.it}
\email{$$m.carriero@istitutotumori.na.it}
\email{$$roberto.natalini@cnr.it}


%\subjclass[2010]{ 82C22, 34D05, 92C17}

\begin{abstract}
Experiments of cell migration and chemotaxis assays have been classically performed in the so-called Boyden Chambers. A recent technology, \emph{xCELLigence} Real Time Cell Analysis, is now allowing to monitor the cell migration in real time. This  technology measures impedance changes caused by the gradual increase of electrode surface occupation by cells during the course of time and provide a Cell Index which is proportional to cellular morphology, spreading, ruffling and adhesion quality as well as cell number. In this paper we propose a macroscopic mathematical model, based on \emph{advection-reaction-diffusion} partial differential equations, describing the cell migration assay using the real-time technology. We carried out numerical simulations to compare simulated model dynamics with data of observed biological experiments on three different cell lines and in two experimental settings: absence of chemotactic signals (basal migration) and presence of a chemoattractant. Overall we conclude that our minimal mathematical model is able to describe the phenomenon in the real time scale and numerical results show a good agreement with the experimental evidences.
\end{abstract}
\maketitle
\numberwithin{equation}{section}
\numberwithin{figure}{section}
\section{Introduction}
Despite significant progress regarding potential therapeutic targets aimed at improving survival, patients affected by solid tumours frequently die for systemic spread of the disease to distant sides. Indeed, when cancer cells acquire the ability to separate and move away from the primary tumour mass, migrate through the surrounding tissue, and enter the lymphatic system and/or blood circulation, the prognosis becomes poor. Therefore, the control of cell motility is a new and attractive approach for the clinical management of metastatic patients. The quantitative assessment of tumour cell migration ability for each patient could provide a new potential parameter predictive of patient outcomes in the future.
\par To metastasise, tumour cells have to early acquire the ability to  move and respond to motogen gradients \citep{wirtz2011}. Cell migration  is a spatially and temporally coordinated multistep process that orchestrates physiological processes such as embryonic morphogenesis, tissue repair and regeneration, and immune-cell trafficking \citep{friedl2000}. When cell migration is deregulated, it contributes to numerous disorders including  tumour metastasis \citep{pantel2004,ridley2003}. Due to its important role in regulating physiological and pathological events, methods aimed to examine cell migration may be very useful and important for a wide range of biomedical research such as cancer biology, immunology, vascular biology, and developmental biology. Migrating cells respond to a plethora of mitogen stimuli, and serum (as mixture of growth factors, cytokines and chemokines) is a major source of chemoattractants. These chemoattractants, through the interaction with their cognate receptors allow cells to acquire a polarized morphology with the extension of adhesive protrusions \citep{ridley2003}. This is followed by the attachment of the protrusion to the substratum at the cell front, the translocation of the cell body and, finally, the detachment of the trailing end of the cell from the substratum \citep{lauffenburger1996,mellado2015}. Such a complex process requires the coupling of extracellular signals with the internal signalling machinery that controls cytoskeleton dynamics \citep{ridley2011}. 
\par The most widely used technique to study cell motility \emph{in vitro} is the Boyden chamber assay in which cells placed in the upper compartment of the chamber are allowed to migrate through a microporous membrane into the lower compartment, in which chemotactic agents are present; after an appropriate incubation time, the membrane between the two compartments is fixed, stained, and the number of cells that have migrated to the lower side of the membrane is determined \citep{zigmond1973}. The subjective nature of measurements and the inability to assess cell motility along the time are the major limitations of this assay. 
\par Current molecular studies are providing a more global physicochemical picture of cell locomotion in which the role of spatial and temporal components of the process are detailed \citep{parsons2010}. Recently, to overcome the manual and highly subjective nature of measurements, accelerate analysis and translate conventional Boyden chamber assay into an automated, quantitative high-throughput system, ACEA Biosciences developed the \emph{xCELLigence} Real Time Cell Analysis (RTCA) technology  able to automatically monitor cell  motility in real-time without the incorporation of labels. The \emph{xCELLigence} RTCA technology measures impedance changes in a meshwork of interdigitated gold microelectrodes located at the bottom side of a microporous membrane (CIM-plate). These changes are caused by the gradual increase of electrode surface occupation by migrating cells during the course of time and provide an index of cell migration. The relative electrical changes during a measurement are displayed by \emph{xCELLigence} software as a unit less parameter termed Cell Index, which is calculated as a relative change in actual impedance divided by a previously registered background value. This method of quantitation is directly proportional to cellular morphology, spreading, ruffling and adhesion quality as well as cell number \citep{ke2011,atienza2006}. To reach a quantitative understanding of the mechanisms underlying these processes, concepts and methods from mathematics and physics can be extremely valuable, as we will see in the following. 
%Thus, we have applied a theoretical analysis to describe cellular motility events. First, we analysed  basal (absence of chemotactic gradient) and directional (presence of serum as a source of chemotactic agents) cell migration of three different cell lines by the \emph{xCELLigence} Real Time Cell Analysis (RTCA) technology. These cell lines have been previously characterized  for their migration ability by us and used as models of three different tumour types \citep{caputo2014,bifulco2012,bifulco2011}. Then we performed numerical simulations to compare the model dynamics using experimental raw data obtained by the \emph{xCELLigence} RTCA in absence or presence of chemotactic gradient. 
\par In general, mathematical models can be very useful to modelize a wide variety of biological systems including cell dynamics and cancer \citep{altrock2015,preziosi-tosin2009,vicsek-collective-cell,dico-zebrafish,dico-flock,dico-conference,dico-cardio}. In particular, the development of quantitative predictive models, based on biological evidence, whose parameters are calibrated on biological data, can help in saving time and resources when designing novel experiments. Moreover, even though a mathematical model is not aimed to replace a real experiment, it can represent a guide to interpret acquired biological data and investigate new insights. In relation to in vitro assays in tumour chamber some mathematical model have been already proposed in the scientific literature, mainly focused on cell invasion experiments. In such context, the invasive ability of the cells is measured by the placement of a coating of extra-cellular matrix proteins on top of the porous membrane. In the papers by \citet{kim2009JMB} and \citet{kim2009Bull} a continuous model, based on partial differential equations (PDEs), was proposed in relation to a Boyden like cell invasion experiment. Then, in \citet{eisenberg2011} the authors proposed a similar model to  
investigate some modulating factors of the cancer cell invasion, making also use of a real-time impedance-based in vitro technology.
\par In this paper, first we analysed  basal (absence of chemotactic gradient) and directional (presence of serum as a source of chemotactic agents) cell migration of three different cell lines by the \emph{xCELLigence} cell analyser: Melanoma A375, fibrosarcoma HT1080, and chondrosarcoma Sarc cell lines have been previously characterized for their migration ability by us and used as models of three different tumour types \citep{caputo2014,bifulco2012,bifulco2011}. Then, we apply a theoretical analysis to describe cellular motility events, and we propose a mathematical model for the cell migration assay using the \emph{xCELLigence} Real Time Cell Analysis (RTCA) technology. The proposed macroscopic model, based on advection-reaction-diffusion equations, adapts and extends the mathematical models in the aforesaid cited papers to the specific in vitro experiment in our analysis (see Section \ref{sec:conclusion}). We calibrated model parameters using real data, as well as information available in scientific and modellistic literature. With such estimate we carried out numerical simulations to compare simulated behaviour with the experimental data in absence or presence of chemotactic gradient. Our numerical results show a very good concordance with the experimental curves. Finally, we validated the model, simulating different experimental conditions, as the initial cell density, and then comparing numerical curves with data obtained from relative experiments. In this regard recorded experimental data on chondrosarcoma Sarc cell line seem to confirm theoretical results. 
\section{Results}\label{}
\bigskip
\subsection{Basal and directional cell migration of three  different cell lines.}\label{sec:3lines}
\begin{figure}[htbp]
	\centering
		\includegraphics[width=1\textwidth]{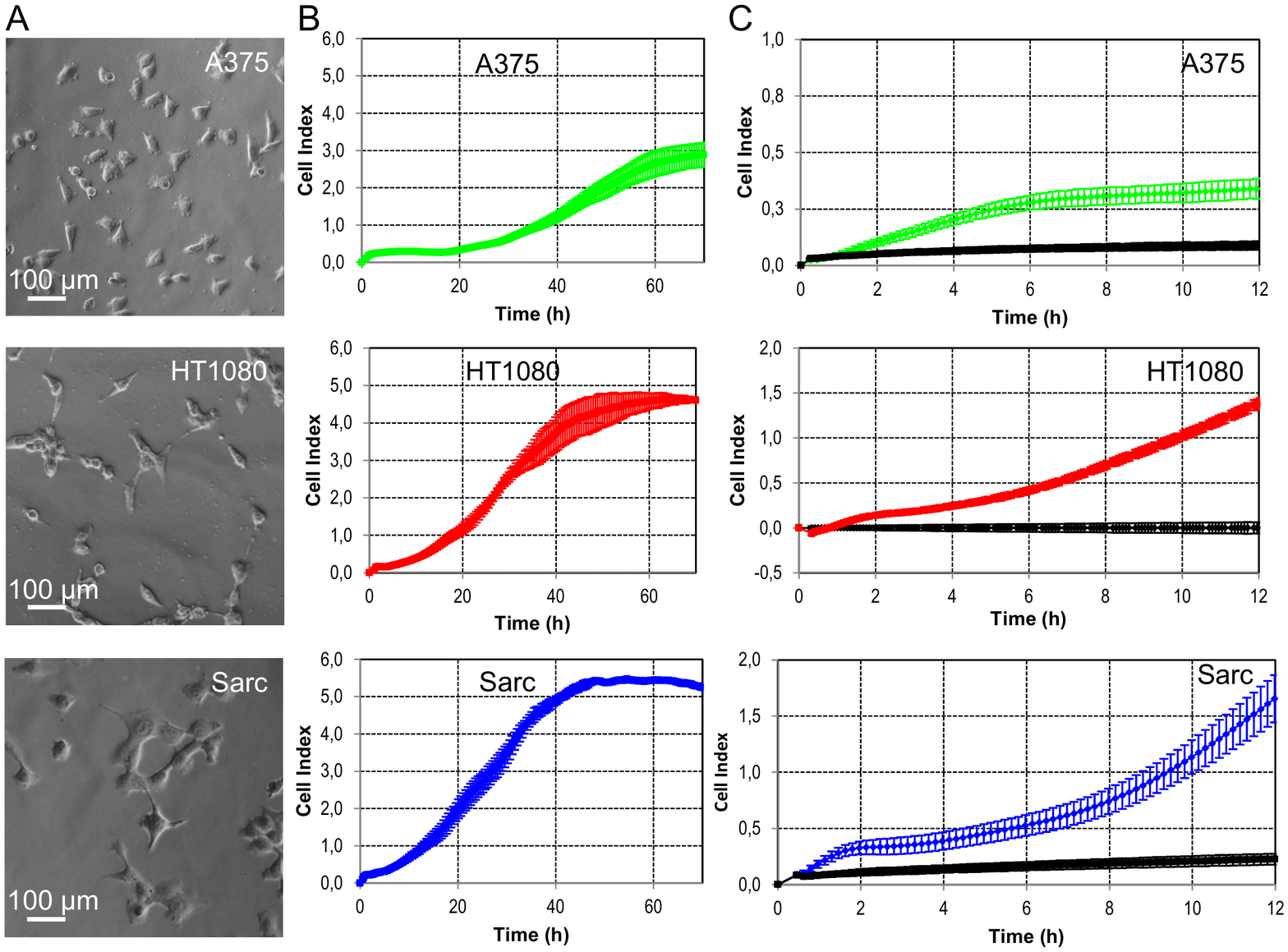}
	\caption{\textbf{Experimental data.} \textbf{A.} Representative images of human melanoma A375,  fibrosarcoma HT1080, or chondrosarcoma Sarc cells analysed by phase contrast microscopy. Original magnifications: 400x. Scale bar: $100\,\mu \mathrm{m}$. \textbf{B.} Time-dependent proliferation of the considered human cell lines. Cells ($2\times10^3$ cells/well) were seeded on E-plates and allowed to grow for 70 h in serum containing medium. The impedance value of each well was automatically monitored by the \emph{xCELLigence} system and expressed as a Cell Index. Data represent mean $\pm$ SD (standard deviation) from a quadruplicate experiment. \textbf{C.} Cell migration  of the indicated human cell lines monitored by the \emph{xCELLigence} system. Cells were seeded on CIM-plates and allowed to migrate towards serum free medium (basal cell migration, black line) or medium plus 10\% FBS. Cell migration  was monitored in real-time for 12 h and expressed as Cell Index. Data represent mean $\pm$ SD from a quadruplicate experiment.}
	\label{panel1}
\end{figure}

\mbox{}\\
For this study we considered three human, neoplastic cell lines which we have previously characterized for their cell migration ability \citep{caputo2014,bifulco2012,bifulco2011}. Melanoma A375,  fibrosarcoma HT1080,  and chondrosarcoma Sarc cell lines (Fig \ref{panel1} panel A) were subjected to both cell proliferation and migration assays using the \emph{xCELLigence}  Real Time Cell Analysis (RTCA) technology. This  technology measures impedance changes in a meshwork of interdigitated gold microelectrodes located at the well bottom (E-plate for proliferation assay) or at the bottom side of a microporous membrane interposed between a lower and an upper compartment (CIM-plate for migration assay). In this way, the impedance-based detection of cell attachment, spreading and proliferation due to the gradual increase of electrode surface occupation may be monitored in real time and expressed as Cell Index.  
To determine the doubling time of A375,  HT1080,  and Sarc cell lines,  cells re-suspended in growth medium were seeded on E-plates and impedance changes were continuously monitored  for 70 h (Fig \ref{panel1}B). Only curves generated by seeding $4\times10^3$ cells/well were considered since those generated by seeding $2\times10^3$ cells/well did not reached a plateau until 90 h. According to their  smaller sized dimension,  A375 cells exhibited a long lasting adhesion/spreading phase and entered the growth phase (proliferation) and then stationary phase (\emph{plateau} phase of growth) due to occupation of all entire microelectrode surface later, as compared to HT1080 and Sarc cells (Figs \ref{panel1}A-B). A375, HT1080 and Sarc cells reached the plateau phase after $\approx$70 h, 65 h, and 50 h, respectively (Fig \ref{panel1}B), and their doubling times calculated from the cell growth curve during the exponential growth were $32.8\pm 1.1$ h, $16.2\pm 0.5$ h and $10.87\pm 0.3$ h, respectively (Fig \ref{figureS1}, Supplementary material).  Since we did not employed cells subjected to cell cycle synchronization, we cannot exclude that, in the presence of serum,  some cell division may occur on the bottom side of filter membranes, thereby affecting Cell Index. Therefore, to minimize the contribution of any cell division,  cell migration experiments were performed for 12 h. 
\par To evaluate cell motility in a system representative of the \emph{in vivo} context, we compared  the ability of A375,  HT1080,  and Sarc cell lines to migrate toward \emph{fetal bovine serum} (FBS) which is a rich source of \emph{growth factor stimuli} and \emph{chemotactic agents}, which signal through binding to their cognate receptors \citep{eisenberg2011}. To this end, cells ($2\times10^4$ cells/well) were seeded on CIM-plates and allowed to migrate toward serum-free medium (basal cell migration) or growth medium, containing 10\% FBS as a source of chemoattractants (directional migration), as described in \citet{carriero2009}. As shown in Fig \ref{panel1}C, all cell lines exhibited a scarce basal cell motility (black lines), as their Cell Indexes did not change significantly along the time. On the other hand, all cell lines were able to respond to serum, although to a different extent. In agreement with their reported high motility \citep{bifulco2012,bifulco2011}, both  fibrosarcoma HT1080 and chondrosarcoma Sarc cells exhibited a comparable, high motility  whereas a low response to FBS was retained by A375 cells (Fig \ref{panel1}C). 
\subsection{The mathematical model}\label{sec:model}
\begin{figure}[htbp]
	\centering
		\includegraphics[width=0.4\textwidth]{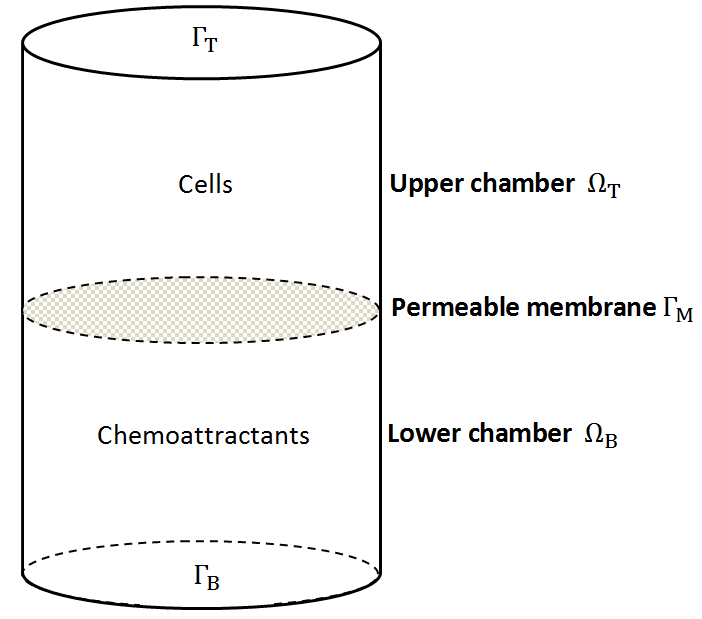}
	\caption{\textbf{Schematic representation of a well of the CIM-plate}. An upper and a lower chamber are separated by a permeable membrane $\Gamma_\text{M}$. In the migration assay in presence of chemoattractant, cells are placed in the upper chamber, and the chemoattractant is added in the lower chamber (directional migration). When measuring the basal migration experiment the well contains only cells (in the upper chamber) and a serum-free medium. In the mathematical formulation the spatial $x\mathrm{-axis}$ is oriented from the top to the bottom.}
	\label{fig:cilindro}
\end{figure}

\mbox{}\\
In our mathematical model we schematise a single well of the CIM-plate used in the experiments as two cylindrical chambers, the upper and the lower chamber respectively, interfaced through the permeable membrane (Fig \ref{fig:cilindro}).
%The micro-electronic biosensor is located on the lower side of the interface. 
The model considers two variables: the cell density and the amount of available FBS, that contains many chemotactic agents. Therefore, we have to take into account the possibility that some chemotactic agents may be degraded, consumed or  internalized during the experiment. Thus, the FBS variable will describe the serum dynamics which includes a possible inactivation of some chemotactic agents. For both cells and the chemical signal we adopt a continuous description. This is justified by the high number of cells (or molecules) involved in the migration assay, typically in the order of $10^5\,\mathrm{cell/cm}^3$.
\par In the following, we first describe the rationale behind the proposed mathematical model, then we provide the explicit formulation in terms of equations. The cell population dynamics is the results of different contributions: diffusion, chemotaxis, spontaneous transport, cell adhesion/spreading. In particular, we consider a diffusion effect of cells in the environmental medium and the chemotactic effect of the FBS that attracts cells toward its higher concentrations. The transport term represents the so called \emph{basal migration}, describing the cell transport through the pores of the permeable membrane, that we experimentally observe also in absence of chemotactic stimuli. The additional term of adhesion/spreading/proliferation represents the increase of Cell Index due to multiple reasons: better adhesion of cells to the biosensor, cell spreading, and possible cell proliferation. Such effects are indistinguishable, since the impedance-based estimate is related to the proportion of biosensor surface in contact with cells. Therefore, a better adherent, or spreaded cell, or duplicated cells produce an analogous increment of the surface contact. In this context, since we are limiting the observation time to 12 h (approximately or below the doubling time, Section \ref{sec:3lines}), the observed effect can be related mostly to adhesion/spreading. The FBS variable is governed by a diffusion effect, coupled with a degradation term due to the chemotactic action (binding) during the cell migration. On the other hand an enzymatic degradation for FBS can be neglected in the considered experimental time range. The permeable membrane is modelled assigning the fluxes of the cell density and of the chemical signal through it, typically proportional to the difference of concentrations on the two sides of the interface.  
\par Now we introduce the general equations of the mathematical model. Let $\Omega$ the domain consisting of the upper ($\Omega_\text{T}$) and lower ($\Omega_\text{B}$) chamber. We indicate with $\Gamma_\text{T}$, $\Gamma_\text{B}$ and $\Gamma_\text{M}$, respectively, the 
boundaries of the upper (\emph{top}) chamber, of the lower (\emph{bottom}) chamber, and the \emph{middle} permeable membrane (see Fig \ref{fig:cilindro}). Then, let $u(\mathbf{x},t)$ the cell density and $\varphi(\mathbf{x},t)$ the FBS concentration, from the above considerations we have:
%
%
%
%\begin{center}
%\fbox{
%\begin{minipage}{0.9 \textwidth}
%\begin{align*}
%\begin{array}{l}
%\text{rate of change of}\\
%\text{cell density}
	%\end{array}	
	%&=\text{diffusion}+\text{chemotaxis effect of FBS}+\text{proliferation/spreading}\\
	%&+\text{spontaneus transport}\\
%\begin{array}{l}
%\text{rate of change of}\\
%\text{FBS signal}
	%\end{array}	
	%&=\text{diffusion}+\text{degradation}\\
	%\begin{array}{l}
%\text{flux of cell density}\\
%\text{on the membrane}
	%\end{array}	
	%&=\text{difference in the cell concentration in upper and lower chamber}\\
		%\begin{array}{l}
%\text{flux of FBS}\\
%\text{on the membrane}
	%\end{array}	
	%&=\text{difference of the FBS concentration in upper and lower chamber}
%\end{align*}
%\end{minipage}
%}
%\end{center}
%
%
%
\begin{align}\label{sys1}
\left\{
\begin{array}{l}
    \overbrace{\partial_t u}^{\text{Cell density rate in time}}=\overbrace{D_u \Delta u}^{\text{Cell diffusion}}-\overbrace{\nabla\cdot \left(u \chi(\varphi) \nabla\varphi \right)}^{\text{Chemotaxis}}-\overbrace{\nabla\cdot \left(\mathbf{V}_{\text{transp}} u \right)}^{\text{Spontaneous transport}}+\overbrace{g(u,\varphi)}^{\text{Cell adhesion/spreading}},\\\\
    \overbrace{\partial_t \varphi}^{\text{Chemoattr. rate in time}}=\overbrace{D_\varphi \Delta \varphi}^{\text{Chemoattr. diffusion}} -\overbrace{\delta u\varphi}^{\text{Binding}},
\end{array}
\right.
\end{align}
where $D_u$, $D_\varphi$, $\delta$ are positive constants, and $\chi(\varphi)$, $g(u,\varphi)$ suitable functions, that will be specified later. On the boundary we assume the following conditions:
\begin{align}
   & \left(D_u\nabla u -u\mathbf{V}_{\text{transp}}\right)\cdot \mathbf{n}=0, \quad &\text{on}\;\Gamma_\text{T},\;\Gamma_\text{B},\label{boud1}\\
   & \nabla \varphi\cdot\mathbf{n}=0, \quad & \text{on}\;\Gamma_\text{T},\;\Gamma_\text{B},\label{boud2}\\
    & \left(D_u\nabla u -u\chi(\varphi)\nabla\varphi-u\mathbf{V}_{\text{transp}}\right)\cdot \mathbf{n}=k_u(u)(u_{\text{B}}-u_{\text{T}}), \quad & \text{on}\;\Gamma_\text{M},\label{boud3}\\
    &		D_\varphi\nabla \varphi\cdot\mathbf{n}=k_\varphi(\varphi_{\text{B}}-\varphi_{\text{T}}), \quad & \text{on}\;\Gamma_\text{M},\label{boud4}
\end{align}
where $\mathbf{n}$ is the downward normal versor, $k_\varphi$ is a constant, while $k_u(u)$ depends on the cell density, and finally $u_{\text{T}}$, $\varphi_{\text{T}}$, $u_{\text{B}}$, $\varphi_{\text{B}}$ are the limit values of $u$ and $\varphi$ on the interface $\Gamma_\text{M}$ from the upper and lower chamber, respectively.
\par Let us now specify the contribution of the different terms in the first and second equation of the proposed system, \eqref{sys1}\textsubscript{1} and \eqref{sys1}\textsubscript{2}. For the diffusion term in \eqref{sys1}\textsubscript{1} we assume a constant diffusion coefficient $D_u$. The chemotaxis term involves a modulating function $\chi(\varphi)$, which takes into account a possible saturation effect for high concentration of chemoattractant. A possible choice is
\begin{align}
\chi(\varphi)&:=\frac{\chi_1\varphi}{\chi_2+\varphi},
\end{align}
with $\chi_1$ and $\chi_2$ positive constants. Similar modelling functions can be found, for example, in \citet{murrayII}. The spontaneous transport of the cell in absence of chemoattractant is modelled as a transport term at velocity $\mathbf{V}_{\text{transp}}$. We can assume a constant velocity in the direction of the vector $\mathbf{n}$ as the limit velocity achieved by the cells in the viscous environment. About the cell adhesion/spreading effect, we consider the function
\begin{align}\label{eq:logistica}
g(u,\varphi):=\alpha_1 u\left(1-\frac{u}{\alpha_3}\right)\frac{\varphi}{\alpha_2+\varphi}\frac{\alpha_2+\bar{\varphi}}{\bar{\varphi}}W(\mathbf{x}),
\end{align}
that establishes a logistic growth for $u$, as it can be deduced from related experiments of proliferation (Sections \ref{sec:3lines}--\ref{metodi:prolif} and Fig \ref{panel1}). The function $W(\mathbf{x})$ is a weight function, which spatially forces the spreading effect as described in the following. Firstly, from the experimental point of view we observe that when cells migrate in the lower chamber, they remain adherent to the bottom side of the membrane. However, for simplicity reasons, in the proposed model cells crossing the membrane are not confined on its lower surface, but they can freely move in the lower chamber. Therefore, to obtain the number of migrated cells, it is necessary to consider the entire lower well, integrating the cell density on it. In our framework, the adhesion/spreading effect involves the cells on the upper face of the membrane and also all cells crossed into the lower chamber. Therefore, a possible choice for the $W$ function, along the $x$-axis, is
\begin{align}\label{eq:W}
	W(x):=
	\left\{
	\begin{array}{ll}
	0,&\quad \text{if }x\leq \bar{x},\\
	\exp\left(-\frac{{(\bar{x}-x_\text{M})}^2}{{(\bar{x}-x_\text{M})}^2-{(x-x_\text{M})}^2}+1\right),&\quad \text{if }\bar{x}<x\leq x_\text{M}, \\
	1,&\quad \text{if }x> x_\text{M},
	\end{array}
	\right.
\end{align}
where $x_\text{M}$ is the position of the central membrane and $\bar{x}$ a suitable constant. In the following we will assume $\bar{x}$ in the order of two cell diameters. The term $\frac{\varphi}{\alpha_2+\varphi}\frac{\alpha_2+\bar{\varphi}}{\bar{\varphi}}$ in Eq \eqref{eq:logistica} considers that FBS serum promotes the cell adhesion/spreading through its growth factors (Section \ref{sec:3lines}), possibly with a saturation effect, while in its absence (for example in the basal migration experiment) such increase in cell density is assumed negligible. The constants $\alpha_1$, $\alpha_3$ can be estimated, for a specific cell line, fitting related proliferation data obtained at concentration of FBS $\varphi=\bar{\varphi}$. The underlining assumption of these estimates is that proliferation assays and migration assay show similar adhesion/spreading rate at least in the earlier times (see Sections \ref{sec:parameters}, \ref{metodi:prolif}). We recall that if the time of observation of the migration assay remains limited in the interval of 12 h, the increment of Cell Index, can be mostly attributed to the adhesion/spreading effect. However, on higher times the contribution of \eqref{eq:logistica} could be able to reproduce also an increase in cell density due to cell proliferation (see also Section \ref{sec:conclusion}).
\par For the FBS signal in \eqref{sys1}\textsubscript{2} we have a diffusion term with constant coefficient $D_\varphi$. Along with this, we consider a serum consuming term proportional to the product between cell density and chemical signal, which takes into account the inactivation of the chemotactic agents due to binding process. Conversely, on the scale of the examined experiment the molecular degradation of the serum can be neglected.  
\par About the boundary conditions, Eqs \eqref{boud1}--\eqref{boud2} represent zero flux for the cell and for the chemoattractant on the top and bottom side of the well, since no mass leaves our domain. In Eqs \eqref{boud3}--\eqref{boud4} we fix Kedem-Katchalsky boundary conditions, meaning that the flux of cells and FBS through the membrane is proportional to the
difference between the concentrations at the top ($u_\text{T}$) and bottom ($u_\text{B}$) sides of the boundary (\citealp{kedem}, and \citealp{quarteroni2002} for a mathematical and modellistic treatment of this conditions). For the $\varphi$ signal we assume a constant transmission coefficient, while for $u$ the transmission term is considered as a function of the cell density. In particular in $k_u(u)$ we assume possible crowding effects on both sides of the interface. A suitable function can be 
\begin{align}\label{eq:ku}
	k_u(u):=\frac{k_{u1}}{1+k_{u2}u_\text{T}+k_{u3}\displaystyle{\left(\int_{\Omega_{\text{B}}} u \,d\mathbf{x}\right)}^p},
\end{align}
where $k_{u1}$, $k_{u2}$, $k_{u3}$  are constants. Notice that \eqref{eq:ku} decreases for increasing cell density on the membrane. In particular, we have two contributions in the denominator: one given by the cell density on the upper side of the interface $\Gamma_M$ ($u_\text{T}$); the other given by a similar contribution on the lower side of $\Gamma_M$, possibly up to the power $p$. As we have observed above, we need to integrate the cell density on the entire lower chamber $\Omega_\text{B}$. Numerical data suggest $p=2$ as a suitable power, which we will assume in the following. All the above considerations are then summarized in the following system of equations:
\small
\begin{align}\label{sys-complete}
\left\{
\begin{array}{ll}
\partial_t u=D_u \Delta u-\nabla\cdot \left(u \displaystyle\frac{\chi_1\varphi}{\chi_2+\varphi} \nabla\varphi \right)-\nabla\cdot \left(\mathbf{V}_{\text{transp}} u \right)+\displaystyle\alpha_1 u\left(1-\frac{u}{\alpha_3}\right) \frac{\varphi}{\alpha_2+\varphi}\frac{\alpha_2+\bar{\varphi}}{\bar{\varphi}}W(\mathbf{x}),&\\\\
\partial_t \varphi=D_\varphi \Delta \varphi-\delta u\varphi,&\\\\
   \left(D_u\nabla u -u\mathbf{V}_{\text{transp}}\right)\cdot \mathbf{n}=0,  &\text{on}\;\Gamma_\text{T},\;\Gamma_\text{B},\\\\
\nabla \varphi \cdot\mathbf{n}=0,  & \text{on}\;\Gamma_\text{T},\;\Gamma_\text{B},\\\\
  \left(D_u\nabla u -u\displaystyle\frac{\chi_1\varphi}{\chi_2+\varphi}\nabla\varphi-u\mathbf{V}_{\text{transp}}\right)\cdot \mathbf{n}=\displaystyle\frac{k_{u1}(u_{\text{B}}-u_{\text{T}})}{1+k_{u2}u_\text{T}+k_{u3}\displaystyle{\left(\int_{\Omega_{\text{B}}} u \,d\mathbf{x}\right)}^2},  & \text{on}\;\Gamma_\text{M},\\\\
  D_\varphi\nabla \varphi \cdot\mathbf{n}=k_\varphi(\varphi_{\text{B}}-\varphi_{\text{T}}),  & \text{on}\;\Gamma_\text{M}.
\end{array}
\right.
\end{align}
\normalsize
Initial concentrations for $u(\mathbf{x},t)$ and $\varphi(\mathbf{x},t)$ will be in the form
\begin{align}
	u(\mathbf{x},0)&=
	\left\{
	\begin{array}{ll}
	u_0,&\quad \text{if }\mathbf{x}\in \Omega_u\subset\Omega_\text{T},\\
	0,&\quad \text{otherwise},
	\end{array}
	\right.\label{eq:u0}\\
	\varphi(\mathbf{x},0)&=
	\left\{
	\begin{array}{ll}
	\varphi_0,&\quad \text{if }\mathbf{x}\in \Omega_\text{B},\\
	0,&\quad \text{otherwise},
	\end{array}
	\right.\label{eq:phi0}
\end{align}
$\Omega_u$ being the portion of the upper chamber, with positive cell density at $t=0$.
%
%
%
%
%
%
%%
%\section{Results}\label{sec:simulation}
%\subsection{Experimental data}
%(Enzo)

%\subsection{Numerical results}
%In this section we present some dynamical tests to simulate our model and compare numerical results on experimental data of migration assays. In order to compare our results with those obtained by xCELLigence Analyzer we express the cell density in term of cell index. 
%%The cell index, experimentally computed, is a measure of the relative change in the electrical impedance in the microelectrical sensor \citep{RTCAsoftware}. 
%The cell index linearly depends on the cell density \citep{xing2005}, and the coefficients of the linear regression can be estimated from an experiment of cell proliferation in which, with the same technology, a known number of cells is put on an impedence-based biosensor and measured their density in time (see Section \ref{sec:parameters}).  
%
%
%
\subsection{Parameter estimation and sensitivity analysis of the mathematical model}\label{sec:parameters}
%\par Experiments on cell migration have been carried out in two typical setting: absence of chemoattractant concentration, and migration in presence of chemotactic stimuli. In the first case the well medium contains only a cell density in the upper chamber. This is useful to study the so called basal migration, which takes into account the spontaneous migration of cells through the membrane due to various concomitant factors, like gravity force and particular morphology of cells. The second case includes the presence of a chemotactic concentration in the bottom chamber.
\mbox{}\\ In our numerical tests we applied the mathematical model 
%in above both cases and 
to three different cell lines: Sarc, HT1080, A375; and two conditions: migration toward chemoattractant (FBS in our case) and basal migration. The second condition corresponds to choose $\chi_1=\alpha_1=0$ in the system \eqref{sys-complete}, so that Eqs \eqref{sys-complete}\textsubscript{1} and \eqref{sys-complete}\textsubscript{2} decouple, and we can simulate only the dynamics of the cell density.  
\par For symmetry reasons, we can simulate a one-dimensional version of system \eqref{sys-complete}, lengthwise the cylindrical domain. Such assumption is in agreement to the impedance-based measurement of the Cell Index performed by the cell analyser, and used to compare numerical data.
\par In order to simulate the dynamic of the model, all parameters have to be chosen. To this purpose we remark that some of them are already available in biological or modellistic literature, while the others have been calibrated on the experimental data. Table \ref{tab-param-dim} summarizes the set of parameters used in our simulations for the different cell lines. For those retrieved from scientific papers we provide the reference in the last column, while for the others, marked as ``data driven'', we specify the experiment (i.e. proliferation, migration, or basal migration) from which we have derived them.
\par In detail, the constants $u_0$, $\varphi_0$, $\bar{\varphi}$ were assigned by the experimental protocol (Sections \ref{metodi:prolif}--\ref{metodi:migration}). The coefficient $D_\varphi$ was set according to \citet{ma2010}. Coefficients related to the cell proliferation, i.e. $\alpha_1$, $\alpha_3$, were obtained, for a specific cell line, from proliferation experiments. This kind of assays was performed in real time on E-plates, using the same technology of the migration ones, for each cell line and at a known FBS concentration $\bar{\varphi}$ (Section \ref{metodi:prolif}). Experimental curves showed a logistic growth in the cell density, and were interpolated to estimate the above mentioned parameters of our interest. Then, the parameters which do not involve chemotactic or growth effects, that are $D_u$, $\mathbf{V}_\text{transp}$, $k_{u1}$, $k_{u2}$, $k_{u3}$, were calibrated on the basal migration curves, fixing in the model $\chi_1=\alpha_1=0$. Finally, $\chi_1$, $\chi_2$, $\alpha_2$, $\delta$, $k_\varphi$, were calibrated consistently with the other parameters, on the migration curves in presence of chemoattractant. 
\par We observe that, as in many mathematical models of biological phenomena, the lack of complete information from the experiments on the parameter values necessarily imposes an uncertainty in the response of the model. To obtain as reliable results as possible, we have studied the influence of the parameters on the behaviour of the model through a \emph{local sensitivity analysis} \citep{saltelli,clarelli}, as described below. Such approach allows us to estimate an influence index between the variation of a parameter and a particular observed output of the model. In our analysis we consider the variation of a single parameter at a time, so interactions among coefficients are 
neglected. This is useful for a first exploration of the parameter space.  
\par Let $p_0$ a parameter value and $\varepsilon$ a small deviation on $p_0$, let $f(p_0)$ an output obtained for the $p_0$ value, we defined the sensitivity index $S$ as the following ratio between relative variations:
\begin{align}\label{eq:sensitivity}
	S:=\frac{\left|f(p_0\pm \varepsilon)-f(p_0)\right|}{f(p_0)}\left(\frac{\varepsilon}{p_0}\right)^{-1}.
\end{align}
Table \ref{tab:sensitivity_index} shows the $S$ value in \eqref{eq:sensitivity} for the parameters that we calibrated on the experimental data. The small deviation $\varepsilon$ was assumed equal to $0.05p_0$, that is a 5\% deviation on the parameter value. The observed output $f$ was the Cell Index at the final time of observation, corresponding to 12 h. Moreover, Table \ref{tab:sensitivity_index} shows also, in the second column, the percentage variation of the examined parameter given by
\begin{align}\label{eq:variazione}
	\Delta f_\text{rel}:=\frac{f(p_0\pm \varepsilon)-f(p_0)}{f(p_0)}100.
\end{align}
\begin{landscape}
\begin{longtable}{@{\extracolsep{\fill}}*{4}{l}}
%\begin{longtable}{\textwidth}{3cm}
\caption{Estimates of initial data, physical and biological parameters. About the model parameters, values were retrieved from scientific literature, or estimated from proliferation or migration assays. For those obtained from migration experiments we used $2\times10^4$ cells/well (Section \ref{sec:parameters}).}
\label{tab-param-dim}\\
\toprule
%\hline
%Parameter & Definition                                                & Estimated value or range (used values)   & Source\\
\multicolumn{1}{l}{\parbox[t][][t]{3cm}{Initial datum\\or parameter}} & \multicolumn{1}{l}{Definition} & 
\multicolumn{1}{l}{Estimated value} & \multicolumn{1}{l}{Source} \\
\midrule
\endfirsthead

% intestazione normale
\multicolumn{2}{l}{\footnotesize\itshape\tablename~\thetable:
continuation from the previous page} \\
\toprule
\parbox[t][][t]{3cm}{Initial datum\\or parameter} & Definition                                                & Estimated value & Source\\
\midrule
\endhead

% piede normale
\midrule
\multicolumn{2}{l}{\footnotesize\itshape\tablename~\thetable:
continuation in the next page} \\
\endfoot

% piede finale
\bottomrule
%\multicolumn{2}{r}{\footnotesize\itshape\tablename~\thetable:
%si conclude dalla pagina precedente} \\
\endlastfoot

%\hline
$u_0$     &\scriptsize \parbox[t][][t]{5.5cm}{initial maximum cell denity in \eqref{eq:u0}}&
\parbox[t][][t]{6cm}{$\approx 30200$, $\approx 45300$, $\approx 60400$ $\mathrm{cell\,cm}^{-1}$ (Sarc)\\
$\approx 30200$ $\mathrm{cell\,cm}^{-1}$ (HT1080, A375)}&\scriptsize \parbox[t][][t]{2.5cm}{Exp. setup: Mat. Meth. \ref{metodi:migration}}\\
%30207, 45311, 60415 (Sarc); 30207 (HT1080, A375)
%\hline
$\varphi_0$     &\scriptsize \parbox[t][][t]{5.5cm}{initial maximum FBS concentration in \eqref{eq:phi0}}&
$18.39\,\mu\mathrm{l\,cm}^{-1}$ & \scriptsize \parbox[t][][t]{2.5cm}{Exp. setup: Mat. Meth. \ref{metodi:migration}}\\
%\hline
$D_u$       & \scriptsize cell diffusion                                      & \parbox[t][][t]{6cm}{$1 \times 10^{-3}$ $\mathrm{cm}^2\,\mathrm{h}^{-1}$ (Sarc)\\
$2.5 \times 10^{-3}$ $\mathrm{cm}^2\,\mathrm{h}^{-1}$ (HT1080)\\
$8 \times 10^{-4}$ $\mathrm{cm}^2\,\mathrm{h}^{-1}$ (A375)}& \scriptsize \parbox[t][][t]{2.5cm}{data driven from\\basal migr. exp. }\\
%\hline 
$D_\varphi$       & \scriptsize FBS diffusion & $3.7 \times 10^{-3}$ $\mathrm{cm}^2\,\mathrm{h}^{-1}$                                 &\scriptsize \citet{ma2010} \\
%\hline
$\chi_1$     & \scriptsize first chemotactic constant&
\parbox[t][][t]{6.5cm}{$3 \times 10^{-3}$ $\mathrm{cm}^3\,\mu \mathrm{l}^{-1}\,\mathrm{h}^{-1}$ (Sarc)\\
$2.5 \times 10^{-3}$ $\mathrm{cm}^3\,\mu \mathrm{l}^{-1}\,\mathrm{h}^{-1}$ (HT1080)\\
$1 \times 10^{-3}$ $\mathrm{cm}^3\,\mu \mathrm{l}^{-1}\,\mathrm{h}^{-1}$ (A375)}                                      
&\scriptsize \parbox[t][][t]{2.5cm}{data driven from\\migr. exp. }\\
%\hline
$\chi_2$     & \scriptsize second chemotactic constant&
\parbox[t][][t]{6cm}{$4.75 \times 10^{-8}$ $\mu\mathrm{l\,cm}^{-1}$\\ (Sarc, HT1080, A375)}                        &\scriptsize \parbox[t][][t]{2.5cm}{data driven from\\migr. exp. }\\
%\hline
$V_\text{transp}$     & \scriptsize transport velocity&
\parbox[t][][t]{6cm}{$9 \times 10^{-3}$ $\mathrm{cm\,h}^{-1}$ (Sarc)\\
$2 \times 10^{-3}$ $\mathrm{cm\,h}^{-1}$ (HT1080)\\
$1.3 \times 10^{-9}$ $\mathrm{cm\,h}^{-1}$ (A375)}                                      
&\scriptsize \parbox[t][][t]{2.5cm}{data driven from\\basal migr. exp. }\\
%\hline
$\alpha_1$     &\scriptsize logistic growth coefficient&
\parbox[t][][t]{6cm}{0.154 $\mathrm{h}^{-1}$ (Sarc)\\
0.135 $\mathrm{h}^{-1}$ (HT1080)\\
0.118 $\mathrm{h}^{-1}$ (A375)}                                      
&\scriptsize \parbox[t][][t]{2.5cm}{data driven from\\prolif. exp. }\\
%\hline
$\alpha_2$     &\scriptsize \parbox[t][][t]{6cm}{dependence on FBS in the logistic growth}&
\parbox[t][][t]{6cm}{$10^{-6}$ $\mu\mathrm{l\,cm}^{-1}$\\ (Sarc, HT1080, A375)}       
&\scriptsize \parbox[t][][t]{2.5cm}{data driven from\\migr. exp.}\\
%\hline
$\alpha_3$     &\scriptsize limit value in the logistic growth&
\parbox[t][][t]{6cm}{$2.08 \times 10^{5}$ $\mathrm{cell\,cm}^{-1}$ (Sarc)\\
$2.26 \times 10^{5}$ $\mathrm{cell\,cm}^{-1}$ (HT1080)\\
$1.04 \times 10^{5}$ $\mathrm{cell\,cm}^{-1}$ (A375)}                                      
&\scriptsize \parbox[t][][t]{2.5cm}{data driven from\\prolif. exp.}\\
%\hline
$\bar{\varphi}$     &\scriptsize \parbox[t][][t]{5.5cm}{FBS concentration in proliferation experiments}&
19.64 $\mu\mathrm{l\,cm}^{-1}$ &\scriptsize \parbox[t][][t]{2.5cm}{Exp. setup: Mat. Meth. \ref{metodi:prolif}}\\
%\hline
$\delta$     &\scriptsize FBS degradation&
\parbox[t][][t]{6cm}{$10^{-8}$ $\mathrm{cm\,h}^{-1}\,\mathrm{cell}^{-1}$ (Sarc, HT1080)\\
$3.5 \times 10^{-5}$ $\mathrm{cm\,h}^{-1}\,\mathrm{cell}^{-1}$ (A375)}                                      
&\scriptsize \parbox[t][][t]{2.5cm}{data driven from\\migr. exp. }\\
%\hline
$k_{u1}$     &\scriptsize cell transmission coefficient on the membrane&
\parbox[t][][t]{6cm}{2 $\mathrm{cm\,h}^{-1}$ (Sarc, HT1080, A375)} &\scriptsize \parbox[t][][t]{2.5cm}{data driven from\\basal migr. exp. }\\
%\hline
$k_{u2}$     &\scriptsize \parbox[t][][t]{6cm}{crowding coefficient on the upper side of the membrane}&
\parbox[t][][t]{6cm}{$1 \times 10^{-5}$ $\mathrm{cm\,cell}^{-1}$ (Sarc)\\
$5 \times 10^{-5}$ $\mathrm{cm\,cell}^{-1}$ (HT1080)\\
$2 \times 10^{-8}$ $\mathrm{cm\,cell}^{-1}$ (A375)} &\scriptsize \parbox[t][][t]{2.5cm}{data driven from\\basal migr. exp. }\\
%\hline
$k_{u3}$     &\scriptsize \parbox[t][][t]{6cm}{crowding coefficient on the lower side of the membrane}&
\parbox[t][][t]{5cm}{$6 \times 10^{-8}$ $\mathrm{cell}^{-2}$ (Sarc, HT1080, A375)} &\scriptsize \parbox[t][][t]{2.5cm}{data driven from\\basal migr. exp. }\\
%\hline
$k_{\varphi}$     &\scriptsize \parbox[t][][t]{6cm}{FBS transmission coefficient on the membrane}& $8.8 \times 10^{-2}$ $\mathrm{cm\,h}^{-1}$ (Sarc, HT1080, A375)&\scriptsize \parbox[t][][t]{2.5cm}{data driven from\\migr. exp.}\\
%\hline
%\hline
\end{longtable}
\end{landscape}
\begin{table}[htbp!]
\caption{Local sensitivity analysis for parameters in Table \ref{tab-param-dim} calibrated from numerical simulations. Second column shows the relative percentage variation as in \eqref{eq:variazione}, choosing as observed output $f$ the Cell Index at the final time of the simulation (12 h), and considering $\varepsilon$ corresponding to a 5\% variation. Third column contains $S$ in \eqref{eq:sensitivity}.}
\label{tab:sensitivity_index}
\begin{center}
\begin{tabular}{l l l}
%\hline\hline
\toprule
 Parameter variation               & Cell Index variation at 12 h& $S$  \\
%\hline\hline
\midrule
$D_u +\varepsilon $    			  &$+0.38\%$&$0.08$\\
$D_u -\varepsilon$    			  &$-0.39\%$&$0.08$\\
$\chi_1+\varepsilon$  			  &$+0.90\%$&$0.18$\\
$\chi_1-\varepsilon$  			  &$-0.97\%$&$0.19$\\
$\chi_2+\varepsilon$   			 	&$-1.2\times 10^{-4}\%$&$2.3\times 10^{-5}$\\
$\chi_2-\varepsilon$   				&$+1.2\times 10^{-4}\%$&$2.3\times 10^{-5}$		 \\
$V_\text{transp}+\varepsilon$ &$+0.59\%$&$	0.12$ \\
$V_\text{transp}-\varepsilon$ &$-0.59\%$&$0.12$	\\
$\alpha_2+\varepsilon$  			&$< 10^{-5}\%$&$< 10^{-5}$ \\
$\alpha_2-\varepsilon$  			&$< 10^{-5}\%$&$< 10^{-5}$ \\
$\delta+\varepsilon$     			&$+7.9\times 10^{-3}\%$&$1.6\times 10^{-3}$\\
$\delta-\varepsilon$     			&$-8\times 10^{-3}\%$&$1.6\times 10^{-3}$\\
$k_{u1}+\varepsilon$     			&$+5.2\times 10^{-2}\%$&$1\times 10^{-3}$\\
$k_{u1}-\varepsilon$     	    &$-5.7\times 10^{-2}\%$&$1.1\times 10^{-3}$\\
$k_{u2}+\varepsilon$     			&$-7.5\times 10^{-5}\%$&$1.5\times 10^{-5}$\\
$k_{u2}-\varepsilon$     			&$+7.5\times 10^{-5}\%$&$1.5\times 10^{-5}$ \\
$k_{u3}+\varepsilon$     			&$-5.2\times 10^{-2}\%$&$1\times 10^{-2}$\\
$k_{u3}-\varepsilon$     			&$5.3\times 10^{-2}\%$&$1\times 10^{-2}$\\
$k_{\varphi}+\varepsilon$			&$+0.15\%$&$0.03$ \\
$k_{\varphi}-\varepsilon$			&$-0.17\%$&$0.03$    \\
\midrule
\end{tabular}
\end{center}
\end{table}
\subsection{Numerical simulations on basal and directional cell migration of three different cell lines}\label{sec:tests}
\mbox{}
\\ In this section we show the performance of our dynamical model in describing experimental results. In order to compare our numerical data with those obtained by \emph{xCELLigence} analyser we express the cell density in term of Cell Index. The Cell Index linearly depends on the cell density \citep{xing2005}, and the coefficients of this linear dependence can be estimated from an experiment of cell proliferation, in which a known number of cells is placed on an impedance-based biosensor and their Cell Index measured in time (Section \ref{metodi:prolif}). 
\par In our simulations we chose the domain $\Omega=[0,1.8]$ (cm) according to the well height in the used CIM-plate, which permeable membrane is placed in the middle, at $x=0.9$ cm \citep{cim-plate}. As observed in previous sections, the observation time was fixed at 12 h (Fig \ref{panel1} in Section \ref{sec:3lines}). For the space discretisation, to preserve stability we adopted a non-uniform mesh. In particular, we fixed $\Delta x=10^{-2}$ cm, while in proximity of the membrane we reduced the spatial step to the finer $\Delta x_{\text{f}}=10^{-6}$ cm. The time step was chosen as the maximum value able to ensure stability and non-negativity of the solution, that is $\Delta t=10^{-3}$ h (see Section \ref{sec:numeric} for further details). 
\par In all numerical tests, the parameters of the model, estimated as described in Section \ref{sec:parameters}, were chosen according to Table \ref{tab-param-dim}. Dynamical simulations were compared with the relative experimental curves computed as described below. For each cell line at least three independent experiments were available. Each experiment was performed in quadruplicate on the same CIM-plate, and the \emph{xCELLigence} data were recorded as mean value (Section \ref{metodi:migration}). We consider as resulting experimental curve for each cell line, the average of the independent replicates (see Fig \ref{figureS2} for full raw data).
\par For each comparison we estimated also the relative MSE error, given by
\begin{align}\label{eq:mse}
\text{MSE}:=\frac{\sum_i^n {(\hat{c}_i-c_i)}^2}{\sum_i^n c_i^2},
\end{align}
where $n$ is the number of experimental time steps, and $\hat{c}_i$, $c_i$ are respectively the numerical and the experimental Cell Index. When necessary, $\hat{c}_i$ was interpolated on time steps of $c_i$. In the following we will indicate with $\text{MSE}_{\text{migr}}$ and $\text{MSE}_{\text{basal}}$ respectively the MSE relative to the migration and basal migration simulations.
%\par Finally, we consider a possible 20\% uncertainty in the initial cell concentration, typically due to errors of measurement (REF....), and we compare the obtained curve bundle with that from experiments.
%
%
%\subsubsection{Study cases: Sarc, HT1080, A375 cell lines}
\par
\begin{figure}[htbp!]
\centering
\subfigure[]{\includegraphics[width=0.5\textwidth, angle=0]{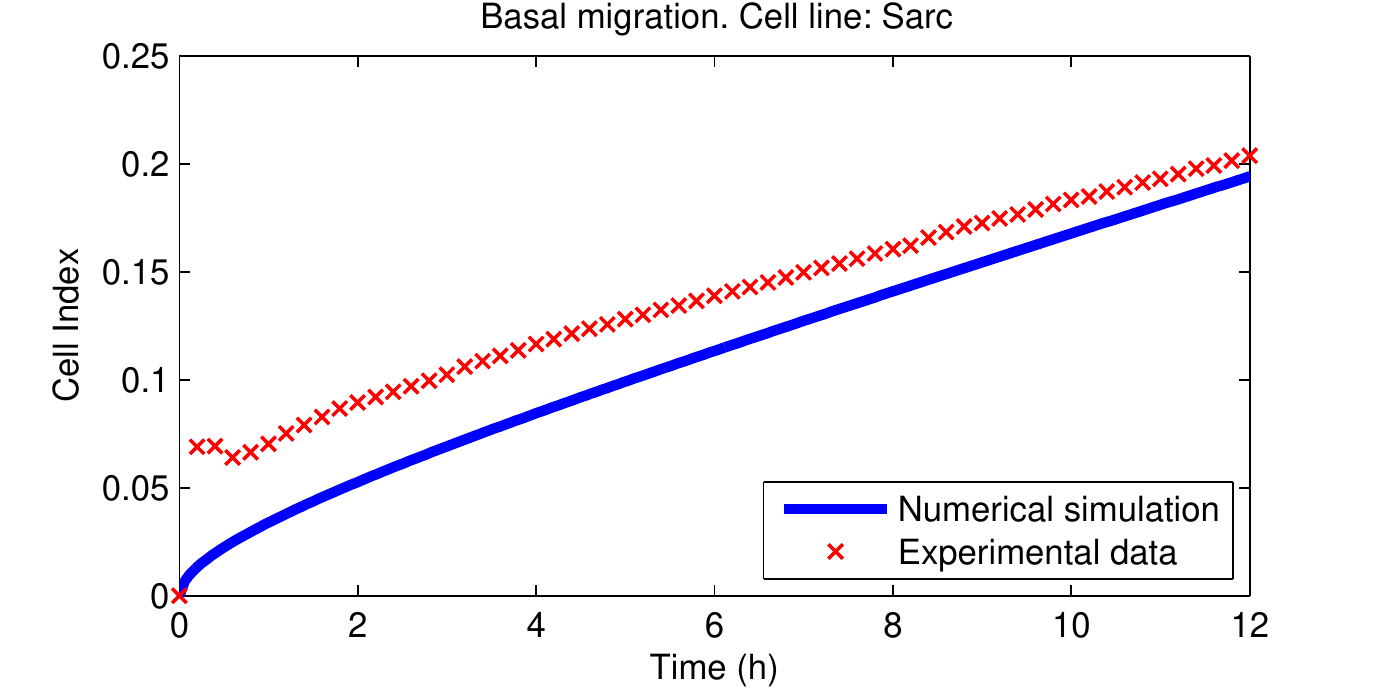}}\hspace{- 0.5 cm}
\subfigure[]{\includegraphics[width=0.5\textwidth, angle=0]{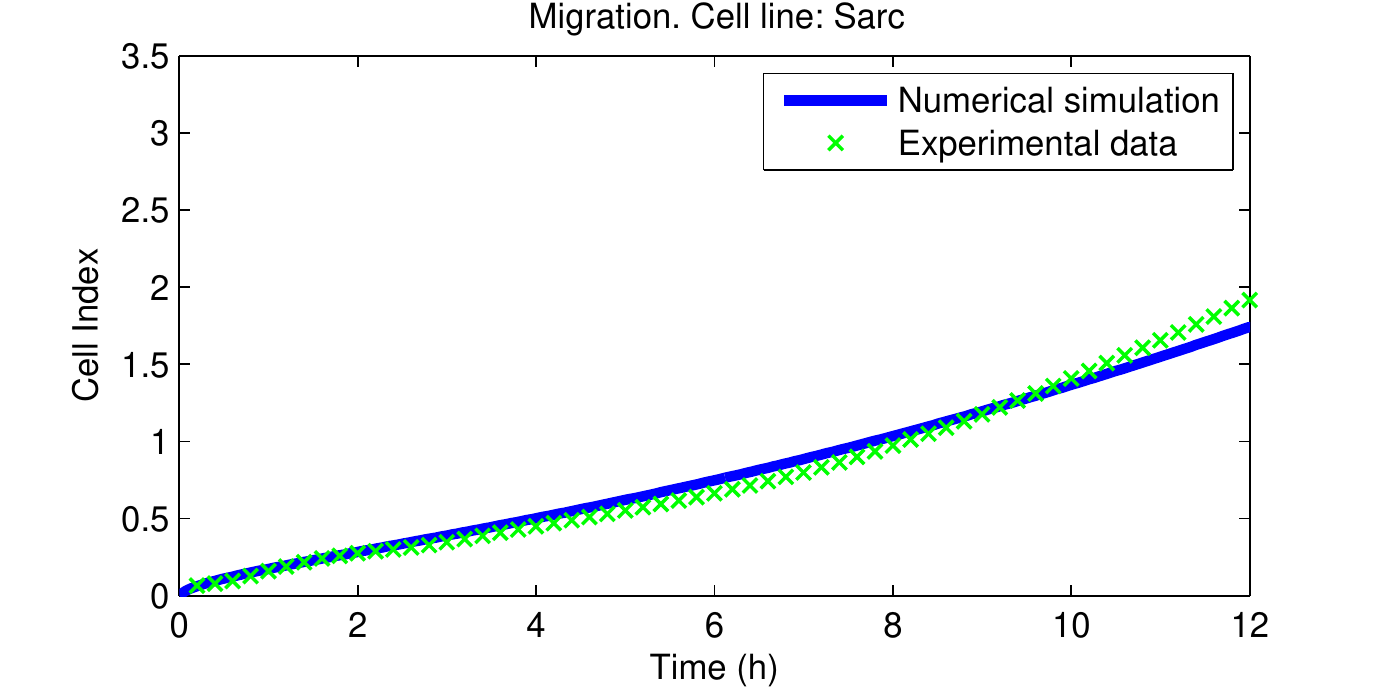}}  \\
\subfigure[]{\includegraphics[width=0.5\textwidth, angle=0]{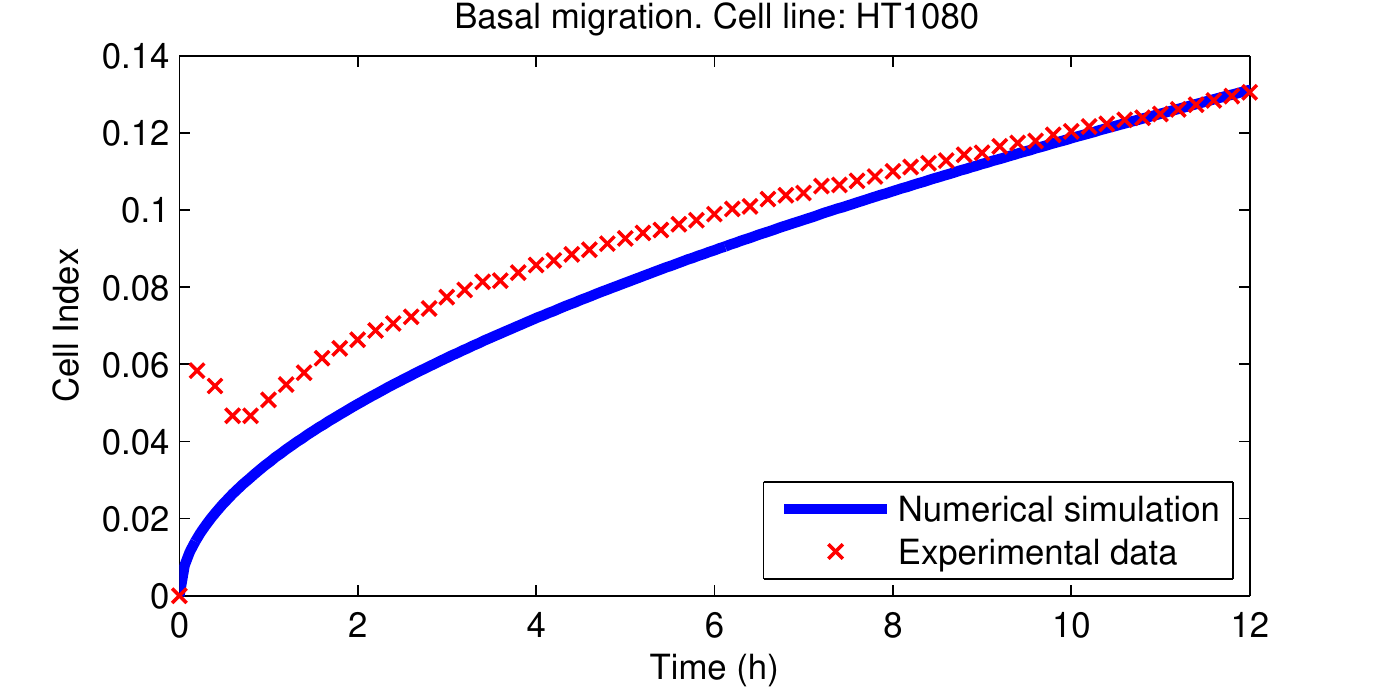}}\hspace{- 0.5 cm}
\subfigure[]{\includegraphics[width=0.5\textwidth, angle=0]{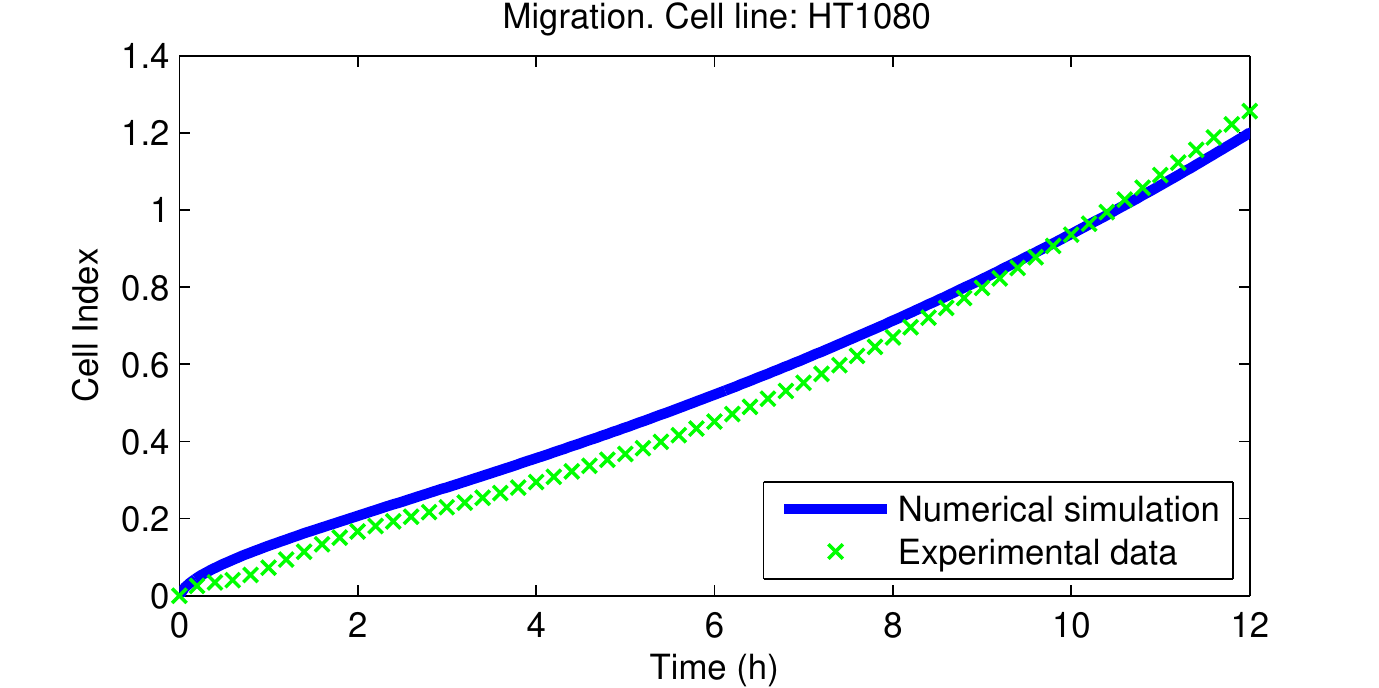}} \\
\subfigure[]{\includegraphics[width=0.5\textwidth, angle=0]{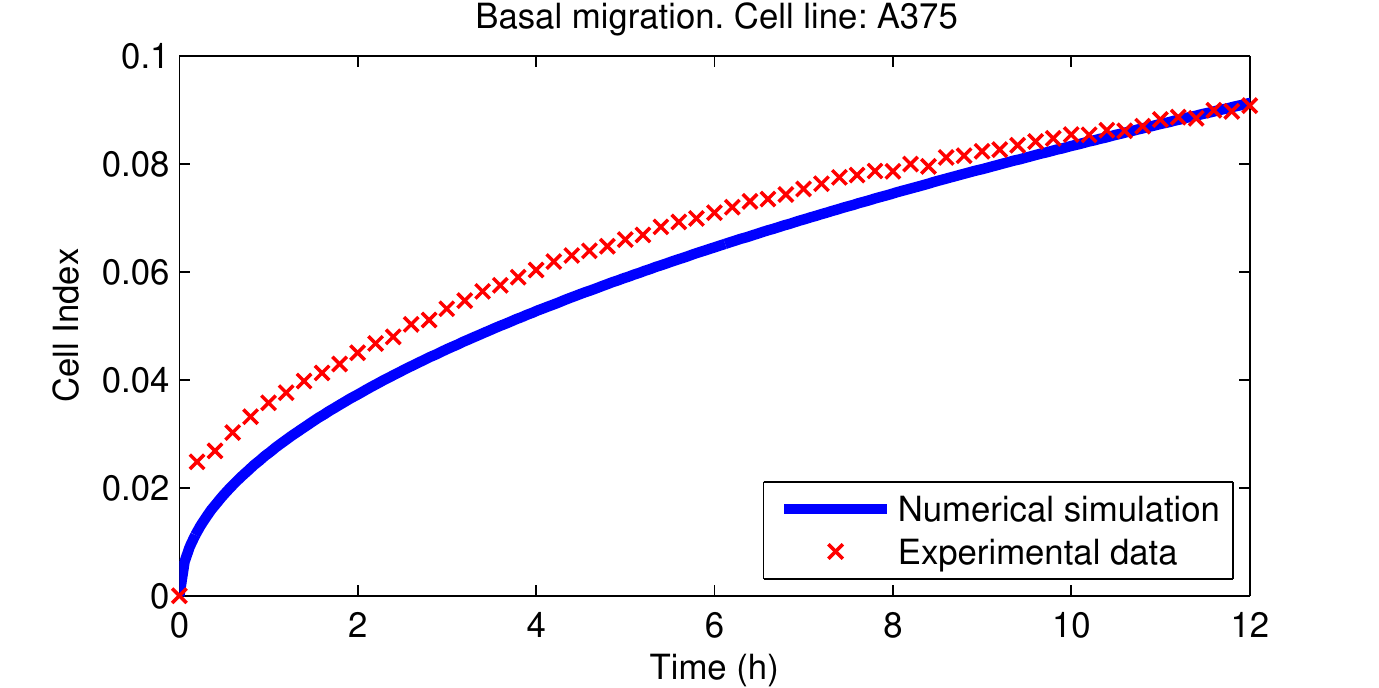}}\hspace{- 0.5 cm}
\subfigure[]{\includegraphics[width=0.5\textwidth, angle=0]{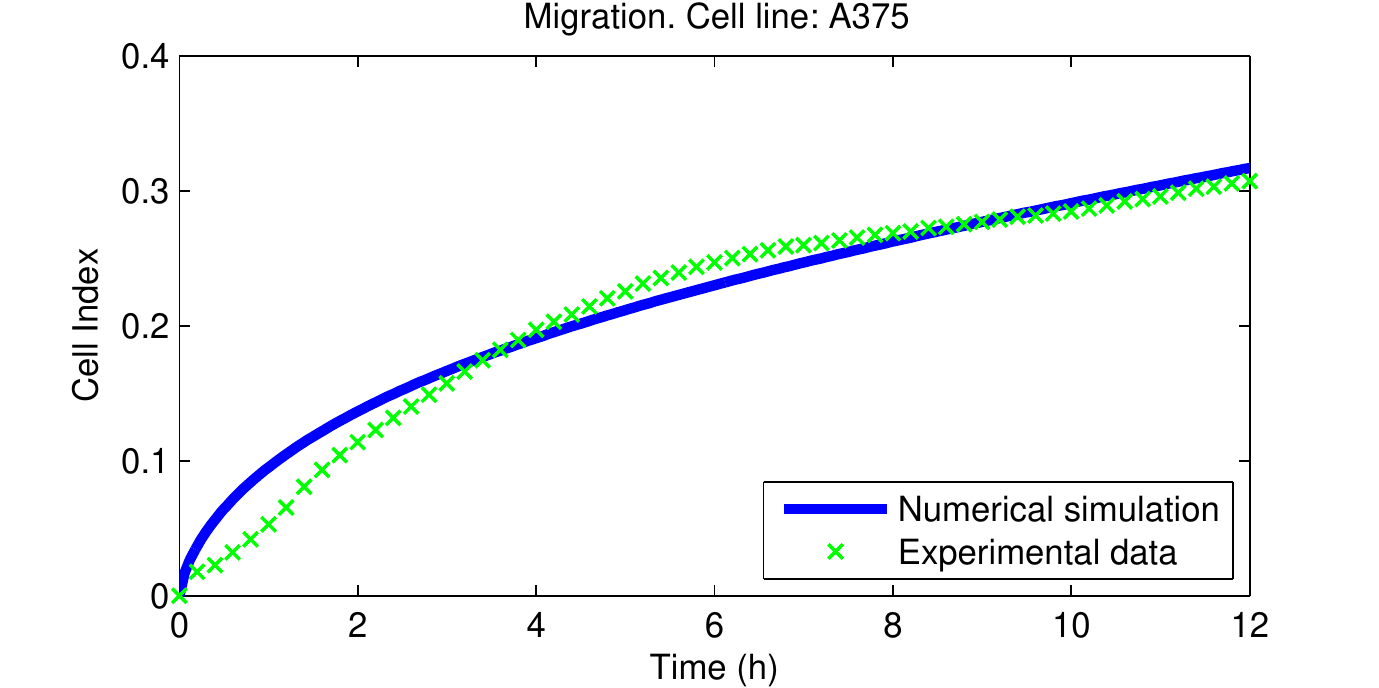}}  
\caption{\textbf{Numerical simulations on Sarc (a)-(b), HT1080 (c)-(d), A375 (e)-(f) cell lines}. For each cell line, panels on the left (a),(c),(e) show the basal migration in absence of chemoattractant. Numerical curves (blue) were compared with experimental data (red). Panels on the right (b),(d),(f) show the migration curves. The simulated values of Cell Index (blue) were compared with experiments (green). Here and in the following figures the experimental curves were obtained as the average of at least three experiments in quadruplicate (Fig \ref{figureS2}). About the MSE value on the Cell Index, defined in \eqref{eq:mse}, we estimated, respectively, the following values: panels (a)-(b) $\text{MSE}_\text{basal}=0.0376$ and $\text{MSE}_\text{migr}=0.0052$; panels (c)-(d) $\text{MSE}_\text{basal}=0.0166$ and $\text{MSE}_\text{migr}=0.0068$; panels (e)-(f) $\text{MSE}_\text{basal}=0.0083$ and $\text{MSE}_\text{migr}=0.0054$.}
\label{fig:sarc119}   
\end{figure}

 Fig \ref{fig:sarc119} shows a numerical simulation of the model \eqref{sys-complete} with parameters fixed as in Table \ref{tab-param-dim} in comparison with the experimental data. Specifically, panel (a) and (b) refer to Sarc cell line, reporting results for basal migration (a) and full system \eqref{sys-complete} (b) respectively. Experimental curves are marked in red for basal migration, and green for chemotactic migration. Both panels display the Cell Index curve versus time. The estimate of the relative MSEs are given by $\text{MSE}_{\text{basal}}=0.0376$ and $\text{MSE}_{\text{migr}}= 0.0052$.
%\bigskip
Figs \ref{fig:sarc119}(c)--(d) refer to HT1080 cell line. In this case we obtain the values $\text{MSE}_{\text{basal}}=0.0166$ and $\text{MSE}_{\text{migr}}= 0.0068$. 
Finally, in Figs \ref{fig:sarc119}(e)--(f) we consider the A375 cell line. For the relative MSE we estimate $\text{MSE}_{\text{basal}}=0.0083$ and $\text{MSE}_{\text{migr}}= 0.0054$. 
\subsection{Confirming the mathematical model on chondrosarcoma Sarc cells}\label{sec:previsione}
\begin{figure}[htbp!]
\centering
\subfigure[]{\includegraphics[width=0.6\textwidth, angle=0]{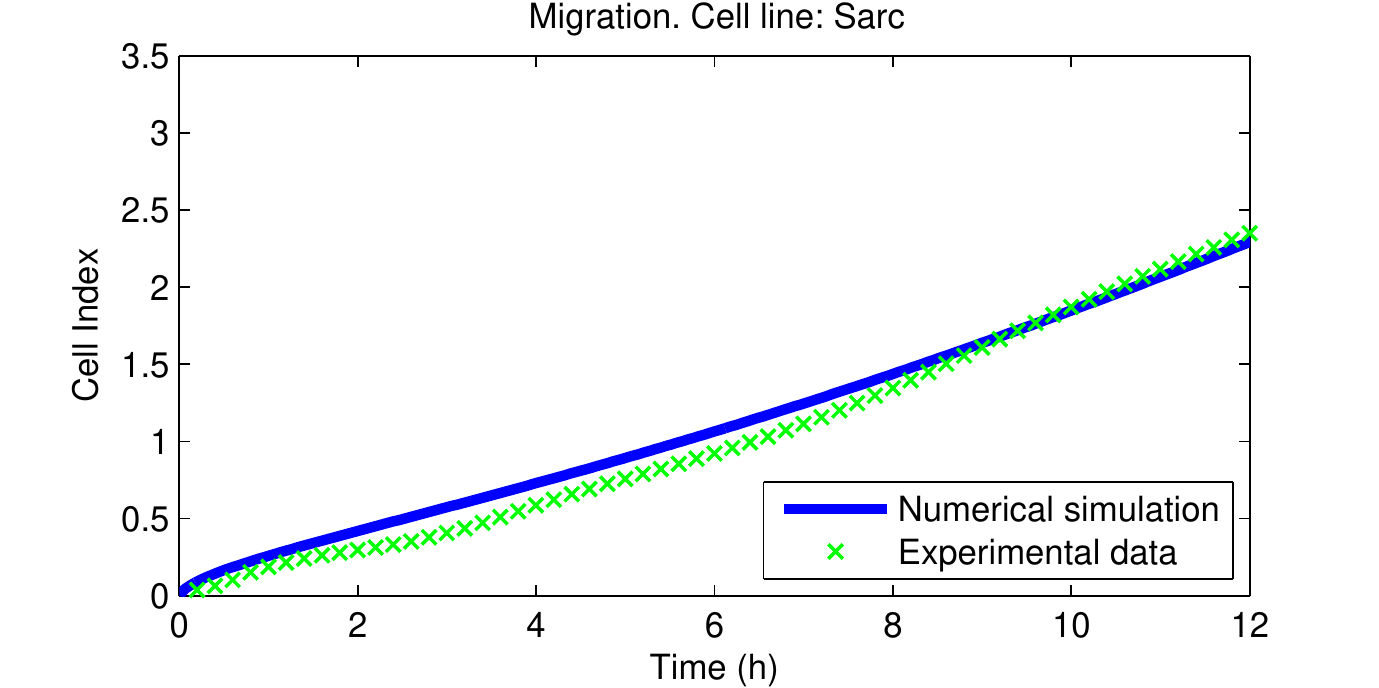}}\\\vspace{1 cm}
\subfigure[]{\includegraphics[width=0.6\textwidth, angle=0]{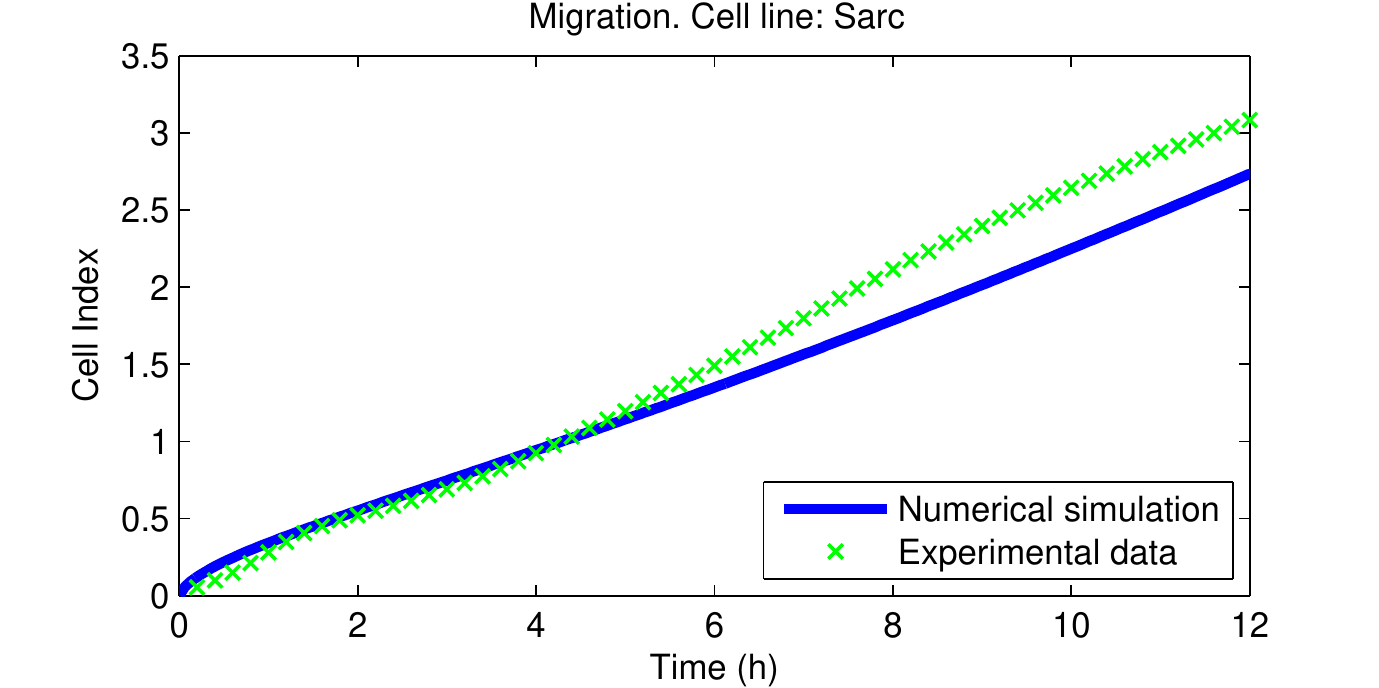}}
\caption{\textbf{Sarc chondrosarcoma cell line. Confirming the mathematical model.} Model \eqref{sys-complete} was simulated with parameters fixed as in Table \ref{tab-param-dim}, obtained with $2\times 10^4$ cells/well, and varying the initial cell density $u_0$. In (a) and (b) numerical data of migration curves were compared with experimental Cell Index respectively in the case of $3\times 10^4$ and $4\times 10^4$ initial cell number. MSE value on the Cell Index was estimated in $\text{MSE}_\text{migr}=0.0077$ and $\text{MSE}_\text{migr}=0.0183$, respectively in (a) and (b).}
\label{fig:sarc1193_30-40}   
\end{figure}

\begin{figure}[htbp]
	\centering
		\includegraphics[width=0.80\textwidth]{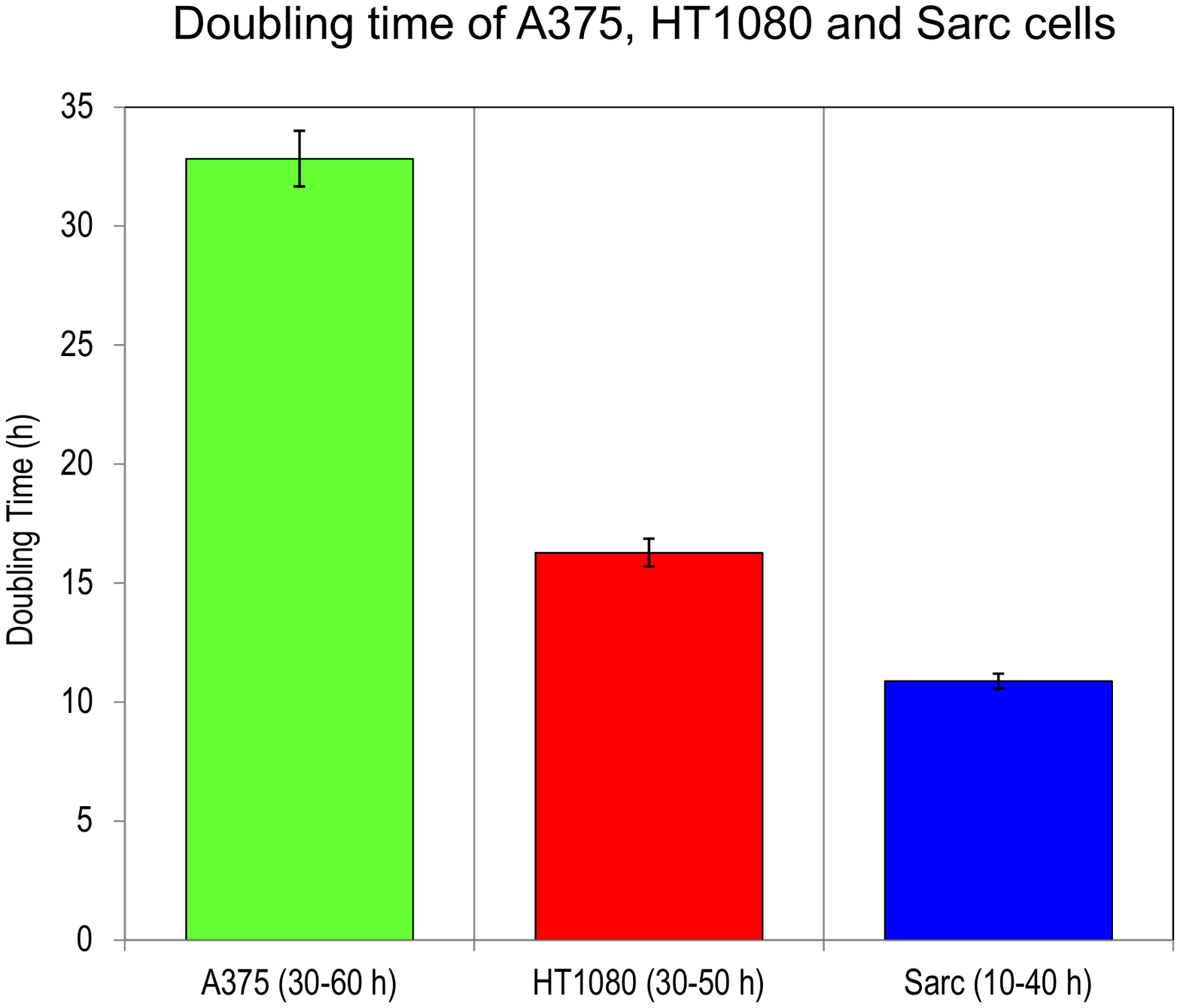}
	\caption{\textbf{Doubling times of Sarc, HT1080, and A375 cell lines.} Cells ($2\times10^3$ cells/well) were seeded on E-plates and allowed to grow for 70 h in serum containing medium. The impedance value of each well was automatically monitored by the \emph{xCELLigence} system and expressed as a Cell Index. Doubling times were calculated, using the \emph{xCELLigence} RTCA software, from the cell growth curves during exponential growth given in round brackets for each cell line. Doubling time is expressed in term of mean value $\pm$ SD (standard deviation) from a quadruplicate experiment.}
	\label{figureS1}
\end{figure}

\begin{figure}[h!]
\centering
\subfigure[]{\includegraphics[width=0.4\textwidth, angle=0]{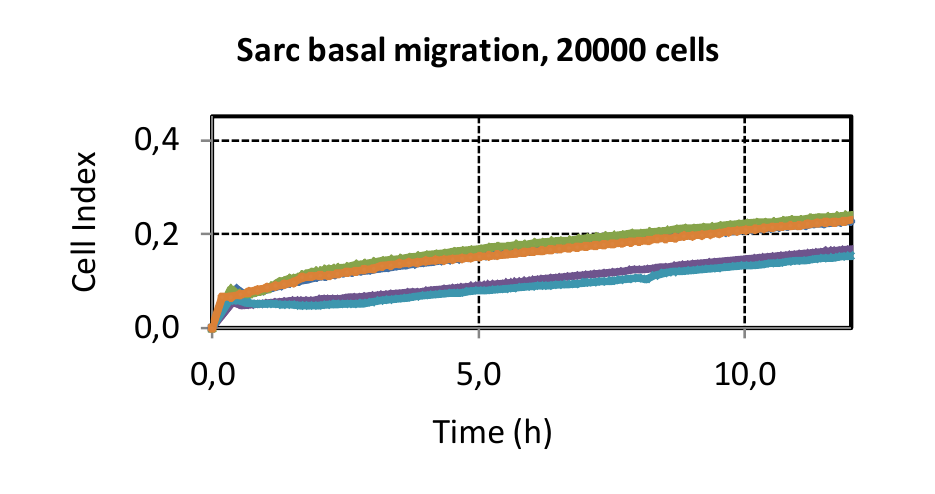}}\hspace{-0.5 cm}
\subfigure[]{\includegraphics[width=0.4\textwidth, angle=0]{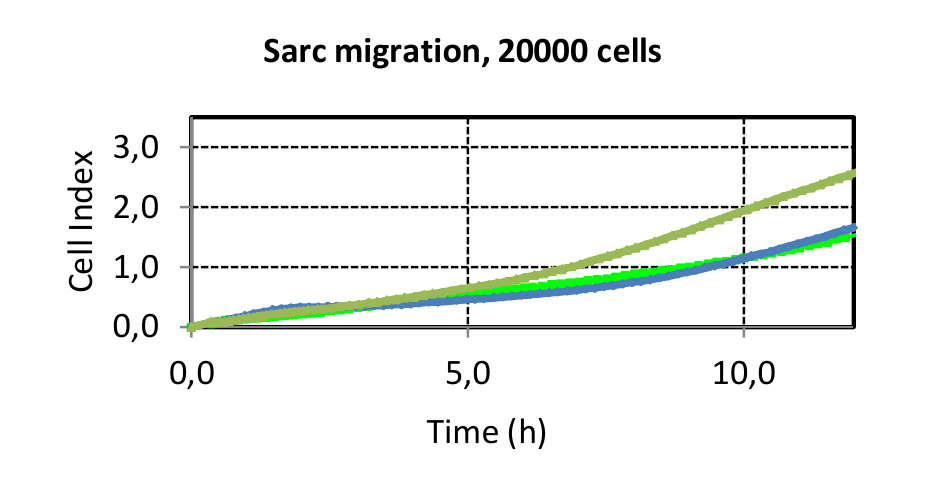}} \\\vspace{- 0.3 cm}
\subfigure[]{\includegraphics[width=0.4\textwidth, angle=0]{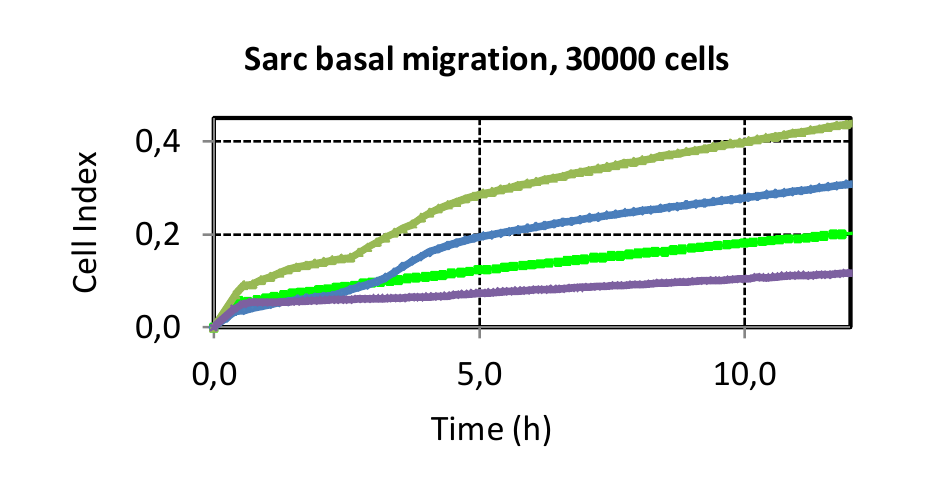}}\hspace{-0.5 cm}
\subfigure[]{\includegraphics[width=0.4\textwidth, angle=0]{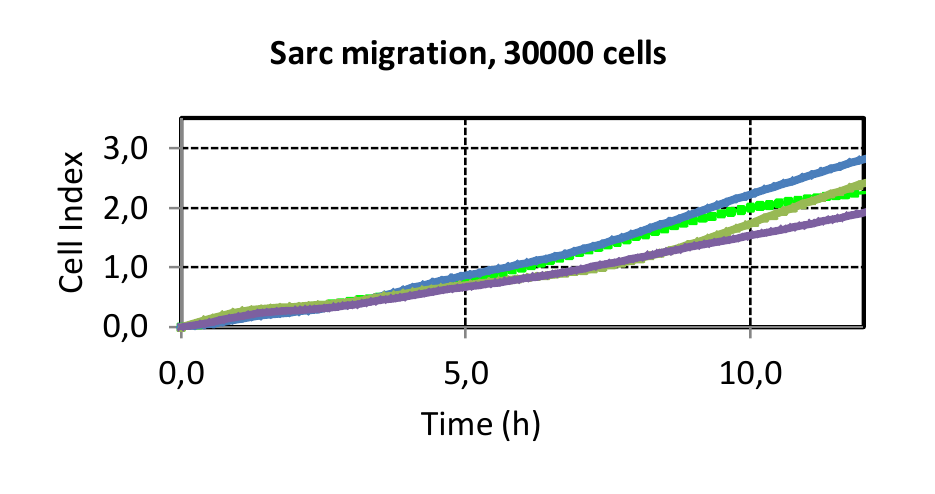}} \\\vspace{- 0.3 cm}
\subfigure[]{\includegraphics[width=0.4\textwidth, angle=0]{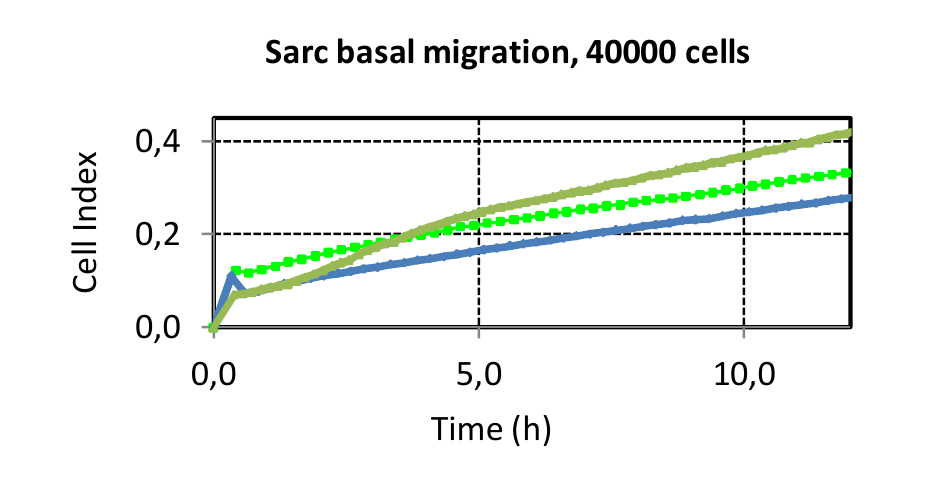}}\hspace{-0.5 cm}
\subfigure[]{\includegraphics[width=0.4\textwidth, angle=0]{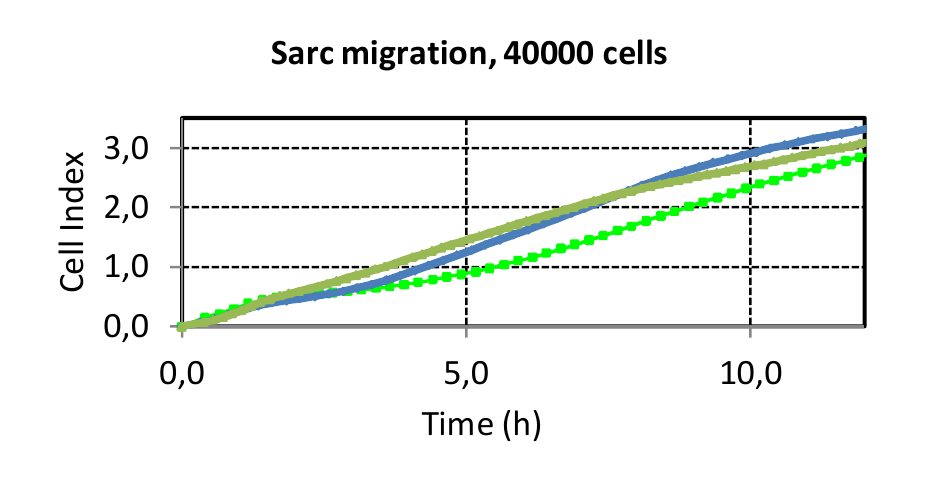}} \\\vspace{- 0.3 cm}
\subfigure[]{\includegraphics[width=0.4\textwidth, angle=0]{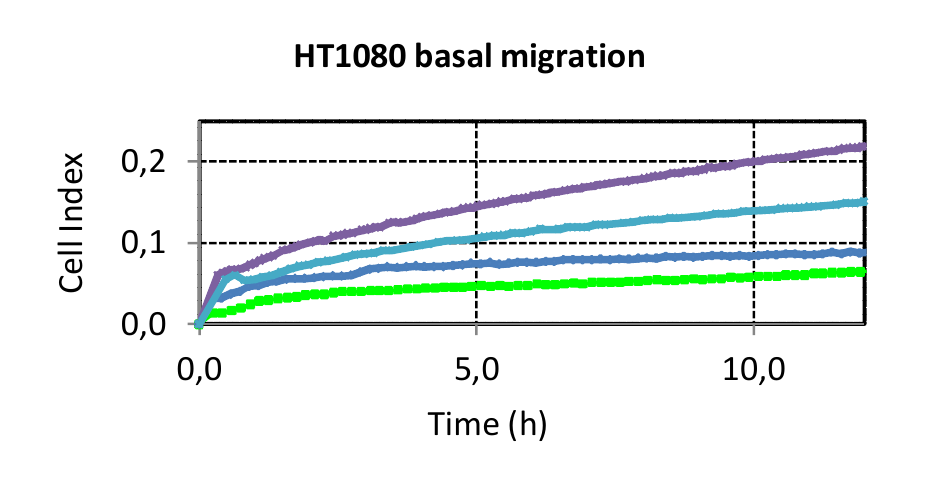}}\hspace{-0.5 cm}
\subfigure[]{\includegraphics[width=0.4\textwidth, angle=0]{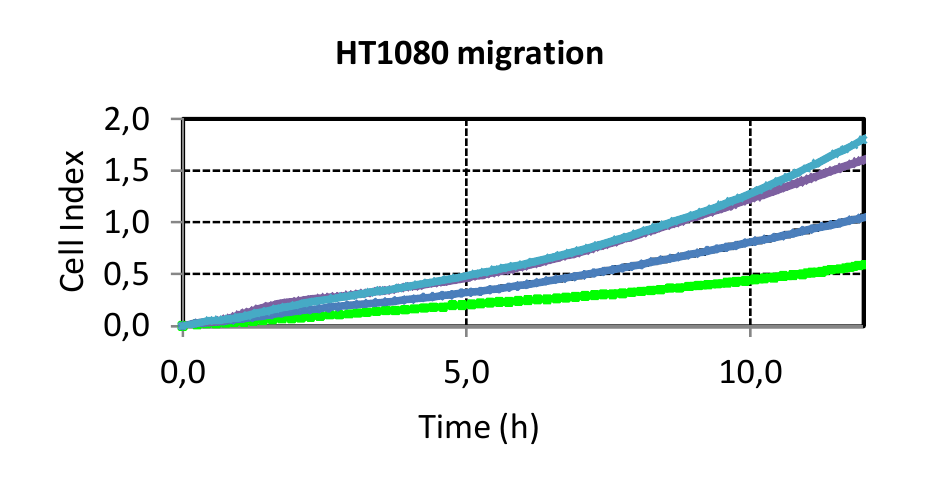}} \\\vspace{- 0.3 cm}
\subfigure[]{\includegraphics[width=0.4\textwidth, angle=0]{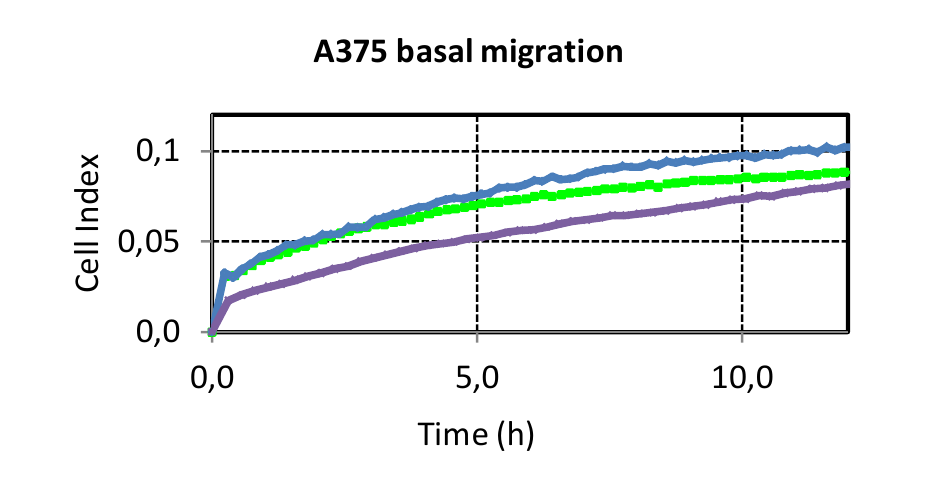}}\hspace{-0.5 cm}
\subfigure[]{\includegraphics[width=0.4\textwidth, angle=0]{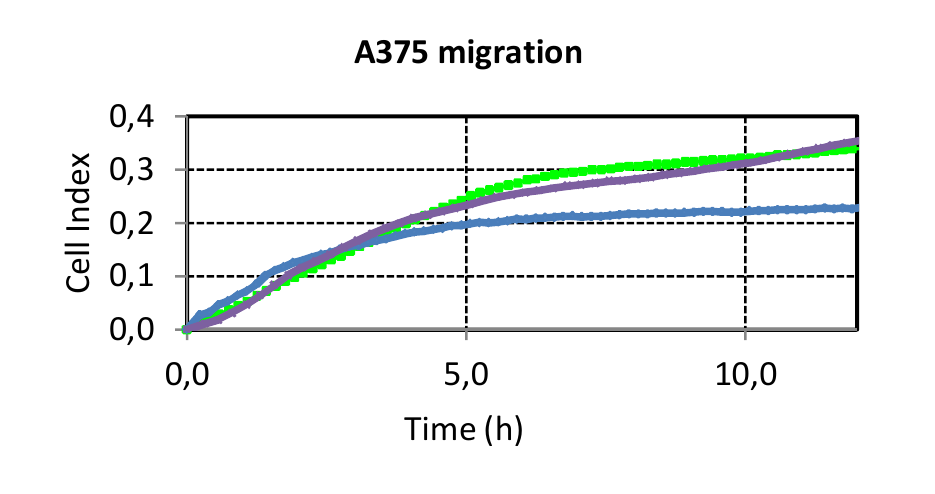}} \vspace{- 0.3 cm}
\caption{\textbf{Cell Index data recorded by \emph{xCELLigence} of the different experiments in our study.} Panels (a),(c),(e),(g),(i) describe the basal migration, (b),(d),(f),(h),(j) the migration in presence of FBS. In each panel the curves represent an independent experiment carried out in quadruplicated and averaged. The observed curves in Figs \ref{fig:sarc119}--\ref{fig:sarc1193_30-40} are obtained as the average of the curves showed here.}
\label{figureS2}   
\end{figure}

\mbox{}\\
In Section \ref{sec:tests} we have shown that, after a suitable parameter calibration, the proposed mathematical model was able to describe the cell migration of three different cell lines, with a very good concordance with the experimental data. Here we investigated the model capability to make predictions about new experiments. To this aim, we used our model to predict the Cell Index on Sarc cell lines performed with different numerosities of cells. 
%Clearly, in principle, other different experimental conditions can be considered, as the serum initial concentration, however the initial cell number represents a simple condition to be experimentally tested. 
Therefore, we applied our mathematical model, using the parameters estimated on the Sarc cell line in the case of $2\times 10^4$ cells in migration (Table \ref{tab-param-dim}), and we estimate the behaviour corresponding to $3\times 10^4$ and $4\times 10^4$ cells/well. Then, in related experiments, cells were seeded at these two different densities on CIM-plates and allowed to migrate towards serum-free medium (basal cell migration) or medium plus 10\% FBS. Cell migration  was monitored in real-time for 12 h as changes in Cell Index. In Fig \ref{fig:sarc1193_30-40} we show the comparison between these numerical curves and Cell Index data obtained by the \emph{xCELLigence} analyser in the case of migration towards chemoattractant. The displayed experimental data represent an average value of three and four different experiments, respectively for the case of $3\times 10^4$ and $4\times 10^4$ cells/well (Fig \ref{figureS2}). In all cases we found a nice agreement with the experimental evidences. In particular, for $3\times 10^4$ and $4\times 10^4$ cells/well, we estimated respectively the relative MSE value as $\text{MSE}_{\text{migr}}=0.0077$, and $\text{MSE}_{\text{migr}}=0.0183$.
%
%
%\begin{figure}[htbp!]
%\centering
%\subfigure[]{\includegraphics[width=0.6\textwidth, angle=0]{figure/ht1080baseline.pdf}}\\\vspace{0 cm}
%\subfigure[]{\includegraphics[width=0.6\textwidth, angle=0]{figure/ht1080migration.pdf}} 
%\caption{\textbf{HT1080}. (a) Basal migration in absence of chemoattractant. Numerical curve (blue) is compared with experiment (red). (b) Migration curve. In the upper panel the concentration of chemoattractant at 12 h. In the bottom panel the simulated values of Cell Index (blue) are compared with experiments (green). MSE value on the Cell Index is estimated in $\text{MSE}_\text{basal}=0.0166$ and $\text{MSE}_\text{migr}=0.0068$.}
%\label{fig:ht1080}   
%\end{figure}
%
%
%\begin{figure}[htbp!]
%\centering
%\subfigure[]{\includegraphics[width=0.6\textwidth, angle=0]{figure/a375baseline.pdf}}\\\vspace{0 cm}
%\subfigure[]{\includegraphics[width=0.6\textwidth, angle=0]{figure/a375migration.pdf}}  
%\caption{\textbf{A375}. (a) Basal migration in absence of chemoattractant. Numerical curve (blue) is compared with experiment (red). (b) Migration curve. In the upper panel the concentration of chemoattractant at 12 h. In the bottom panel the simulated values of Cell Index (blue) are compared with experiments (green). MSE value on the Cell Index is estimated in $\text{MSE}_\text{basal}=0.0083$ and $\text{MSE}_\text{migr}=0.0054$.}
%\label{fig:a375}   
%\end{figure}
%
%
%
%
\clearpage
\section{Discussion and conclusions}\label{sec:conclusion}
Cell migration is a process that offers rich targets for intervention in key pathologic conditions, including cancer.  Indeed, the development of metastases requires the activation of a series of physiological and biochemical processes that govern the migration of tumour cells from the primary tumour site, the invasion through the basement membrane, the entry of metastatic cells into the blood vessels and finally localization to the second site \citep{wirtz2011}. Therefore, targeting cell motility has been increasingly accepted as a new approach for the clinical management of metastatic patients and in the future, quantitative analysis of the motility of tumour cells derived from cancer  patients could provide a new potential parameter predictive of patient outcomes. The recent expansion of mathematical modelling is already contributing to cancer research by helping to elucidate mechanisms of  tumour initiation, progression and metastases as well as intra-tumour heterogeneity, treatment responses and resistance \citep{altrock2015}. Parametrization of cell motility is often difficult given the available experimental model systems. With the advent of high throughput systems, there has been a movement towards the use of a number of cell-based assays useful for studying cell migration. A recent technology, \emph{xCELLigence} RTCA, has been increasingly accepted as a platform for high-throughput determination of cell motility dynamics in real time using micro-electronic biosensors \citep{limame2012}. In this paper we propose a macroscopic mathematical model, based on convection-reaction-diffusion equations, for the cell migration assay. 
\par Previous mathematical models on in vitro Boyden-like assays dealt mainly with the invasion experiment \citep{kim2009JMB,kim2009Bull,eisenberg2011}. Among these, \citet{eisenberg2011} studied cancer cell invasion through a theoretical model compared with real-time impedance-based assays. By contrast, in this paper, we proposed a PDEs model in relation to the cell migration experiment relying on the \emph{xCELLigence} real-time technology. Our model differs from \citet{eisenberg2011}, where it was assumed that the simulated Cell Index is proportional to the fraction of cells that reaches the upper well bottom. The authors assumes that the pores dimension of the permeable membrane are larger enough to allow cells, quite easily, to cross it. On the contrary, cell lines employed in our migration experiments present much larger dimensions than membrane pores ($8\,\mu\mathrm{m}$) (Fig \ref{panel1}, panel A). For this we consider the effective cell crossing through the porous interface, and the simulated Cell Index was computed on the basis of the fraction of cells migrating in the lower chamber. Moreover, our transmission coefficient in the boundary conditions includes also possible crowding effects, being assumed as a decreasing function of the cell density on both the faces of the separating membrane. Finally, to model the basal migration effect, as described in Section \ref{sec:model}, our model considered also a spontaneous transport of cells across the permeable membrane, present even in absence of chemotactic stimuli, that is not considered in \citet{eisenberg2011}. This allowed us to recover experimental data through the basal experiment and to estimate on it some parameters to be included in the full migration model.
%\par Starting from our minimal mathematical model we were able to describe the migration phenomenon in the real time scale, and numerical results showed a nice agreement with the experimental data acquired with \emph{xCELLigence} analyser for three examined cell lines, in both conditions of presence and absence of chemoattractant. Then, we tested the quantitative accuracy of the model to reproduce cell migration in different experimental conditions. On chondrosarcoma Sarc cell line we used the calibrated model to simulate an experiment with two different initial cell densities. Our numerical results were confirmed with a very good approximation by acquired data on related experiments.
\par Numerical simulations has been performed to compare the model dynamics with experimental raw data obtained by the \emph{xCELLigence} RTCA in absence or presence of a chemotactic gradient. We also validate the performance of our model by comparing the results of simulations with other experimental data, on chondrosarcoma Sarc cell line, not used for estimating model parameters. Numerical findings showed a nice agreement with the acquired experimental data. Therefore, overall we can infer that tumour cells migration can be described using mathematical models as a predictable process dependent on biophysical laws and experimental parameters.  
\par Starting from the present paper, some interesting issues can be investigated as future perspectives. Firstly, we could explore the quantitative and qualitative accuracy of the model to simulate different experimental conditions in a migration assay, such as the initial serum concentration, or to test the effects of various chemoattractants. In this regard, it could also be interesting to introduce the action of chemotactic inhibitors on the cell motility. In the future, we will explore the possibility of simulating in silico the ability of inhibitors of cell migration to counteract the motility of primary tumour cells derived from patients affected by solid tumours, in order to design more personalized therapeutic strategies. 
\section{Materials and methods}
%\subsection{Experimental setup}
\subsection{Cell Lines}
Human melanoma A375 cell line  was cultured in RPMI 1640 medium (Lonza, Milan, Italy), supplemented with 3 mM L-glutamine (Invitrogen-Gibco\textregistered/Life Technologies, Monza, Italy), 2\% penicillin/streptomycin and 10\% fetal bovine serum (FBS). Highly mobile human fibrosarcoma HT1080 cell line \citep{bifulco2012} and human chondrosarcoma Sarc cells derived from a chondrosarcoma primary culture \citep{bifulco2011} were cultured  in Dulbecco Modified Eagle Medium (DMEM) supplemented with 10\% fetal bovine serum (FBS), 100 IU/ml penicillin and $50\,\mu\mathrm{g/ml}$ streptomycin. All cells were maintained at $37^{\circ}\mathrm{C}$ in a humidified atmosphere of 5\% CO\textsubscript{2}.
\subsection{Cell Proliferation}\label{metodi:prolif}
Cell proliferation was assessed using the \emph{xCELLigence} RTCA technology as described \citep{yousif2015}. For these experiments, the impedance-based detection of cell attachment, spreading and proliferation was assessed by using E-plates which are provided of microelectrodes attached at the bottom of each well. First, $100\, \mu\mathrm{l}$ of growth medium was added to each well, the  plate was locked at $37^{\circ}\mathrm{C}$ in a humidified atmosphere of 5\% CO\textsubscript{2} and the background impedance was measured. Cells were counted, suspended in in $100\,\mu\mathrm{l}$ growth medium,  seeded ($2\times 10^3$ or $4\times 10^3$ cells/well) and allowed to grow for 70 h. The impedance value of each well was automatically monitored by the \emph{xCELLigence} system and expressed as a Cell Index.
\subsection{Cell Migration}\label{metodi:migration}
Cell migration was monitored using the \emph{xCELLigence} RTCA technology as described \citep{yousif2015}. For these experiments, the impedance-based detection of cell migration was assessed using CIM-plates which are provided of interdigitated gold microelectrodes on bottom side of a microporous membrane (containing randomly distributed $8 \,\mu\mathrm{m}$-pores) interposed between a lower and an upper compartment. Briefly, $160\, \mu\mathrm{l}$ of serum-free medium with/without 10\% FBS and $30\,\mu\mathrm{l}$ of serum-free medium were added to the lower and upper chambers, respectively, prior to lock the plate at $37^{\circ}\mathrm{C}$ in a humidified atmosphere of 5\% CO\textsubscript{2} for 60 minutes (to obtain the equilibrium between the two compartments), according to the manufacturer's guidelines. Then, background signals generated by cell-free media were measured, detached cells were counted, suspended in $100\, \mu\mathrm{l}$ serum-free medium and seeded ($2\times 10^4$, $3\times 10^4$, $4\times 10^4$ cells/well) in the upper chamber. Microelectrodes detect impedance changes which are proportional to the number of migrating cells and are expressed as Cell Index. Cell migration was monitored in real-time for 12 h. Each experiment was performed at least three times in quadruplicate.

\subsection{Numerical methods}\label{sec:numeric}
%\mbox{}\\
The numerical approximation scheme used in the simulation of the model \eqref{sys-complete} employed a finite difference method on a spatial domain $\Omega=[a,b]$, consisting of upper and lower domains $\Omega_\text{T}$, $\Omega_\text{B}$, interfaced through the membrane $\Gamma_\text{M}$. Let $\Delta x$, $\Delta t$ the space and time steps, we defined the grid points $(x_i,t_k)$, where $x_i=i\Delta x$ and $t_k=k\Delta t$. The approximation of a function $f(x,t)$ at the grid point $(x_i,t_k)$ was denoted as $f_i^k$. To ensure non-negativity in the numerical simulations, due to the boundary conditions on the permeable membrane, in its proximity we needed to discretise our equations on a finer spatial mesh $x_j=j\Delta x_\text{f}$, $\Delta x_\text{f}<\Delta x$. 
\par For the diffusion equation \eqref{sys-complete}\textsubscript{2} we applied on the internal nodes a central scheme in space and an implicit scheme in time for the diffusive term, while the reaction term was put in explicit:
\begin{align*}
	\frac{\varphi_{i}^{k+1}-\varphi_{i}^{k}}{\Delta t}=D_{\varphi}\frac{\varphi_{i+1}^{k+1}-2\varphi_{i}^{k+1}+\varphi_{i-1}^{k+1}}{\Delta x^2}-\delta u_i^k\varphi_i^k.
\end{align*}
Similarly on the finer mesh with spatial step $\Delta x_\text{f}$. 
\par For the advection-diffusion equation \eqref{sys-complete}\textsubscript{1}, let 
\begin{align}
	V:=\frac{\chi_1 \varphi}{\chi_2+\varphi}\partial_x\varphi+{V}_{\text{transp}},
\end{align}
we assumed
\begin{align}
	V_i^k&=\chi_i^k\frac{\varphi_{i+1}^{k}-\varphi_{i-1}^{k}}{2\Delta x}+V_{\text{transp}},\\
	\chi_i^k&:=\frac{\chi_1 \varphi_i^k}{\chi_2+\varphi_i^k},\label{chi_ik}
\end{align}
and for the internal nodes we adopted the scheme 
\begin{align*}
	\frac{u_{i}^{k+1}-u_{i}^{k}}{\Delta t}&=D_u\frac{u_{i+1}^{k+1}-2u_{i}^{k+1}+u_{i-1}^{k+1}}{\Delta x^2}-\frac{V_{i+1}^ku_{i+1}^{k}-V_{i-1}^ku_{i-1}^{k}}{2\Delta x}\\
	&+\alpha_1 u_{i}^{k}\left(1-\frac{u_{i}^{k}}{\alpha_3}\right) \frac{\varphi_{i}^{k}}{\alpha_2+\varphi_{i}^{k}}\frac{\alpha_2+\bar{\varphi}}{\bar{\varphi}}W(x_i)+\frac{|V|_{i+1}^ku_{i+1}^{k}-2|V|_{i}^ku_{i}^{k}+|V|_{i-1}^ku_{i-1}^{k}}{2\Delta x},	
\end{align*}
with function $W(x_i)$ defined in \eqref{eq:W}, and where the last term introduced an artificial viscosity in order to preserve scheme stability (see for example \citealp{bracciale2015}).
\par For the boundary conditions \eqref{sys-complete}\textsubscript{3,4} on $\Gamma_\text{T}$ ($x=x_0$) and on $\Gamma_\text{B}$ ($x=x_N$) we used the second order one-sided approximation of the second derivative in the form:
\begin{align*}
	\frac{D_u}{2\Delta x} \left(-3 u^k_0+4 u^k_1-u^k_2\right)-V_{\text{transp}} u_0^k&=0,\\
	\frac{1}{2\Delta x} \left(-3 \varphi^k_0+4 \varphi^k_1-\varphi^k_2\right)&=0,\\
	\frac{D_u}{2\Delta x} \left(3 u^k_N-4 u^k_{N-1}+u^k_{N-2}\right)-V_{\text{transp}} u_N^k&=0,\\
	\frac{1}{2\Delta x} \left(3 \varphi^k_N-4 \varphi^k_{N-1}+\varphi^k_{N-2}\right)&=0.
\end{align*}
On the boundary $\Gamma_\text{M}$ ($x=x_\text{M}$), let $u_\text{T}$, $u_\text{B}$ the variable $u$ in $\Omega_\text{T}$, $\Omega_\text{B}$ respectively. From \eqref{sys-complete}\textsubscript{5,6}, for $u$ we employed    
\begin{align*}
	\frac{D_u}{2\Delta x} \left(3 u^k_\text{T,M} -4 u^k_\text{T,M-1}+u^k_\text{T,M-2}\right)-u^k_\text{T,M} \chi_i^k\frac{k_{\varphi}}{D_\varphi} (\varphi^k_\text{B,M}-\varphi^k_\text{T,M})&-V_{\text{transp}}^k u^k_\text{T,M}\\
	&=(k_u)_i^k(u^k_\text{B,M}-u^k_\text{B,M}),\\
	\frac{D_u}{2\Delta x} \left(-3 u^k_\text{B,M} +4 u^k_\text{B,M+1}-u^k_\text{B,M+2}\right)-u^k_\text{B,M} \chi_i^k\frac{k_{\varphi}}{D_\varphi} (\varphi^k_\text{B,M}-\varphi^k_\text{T,M})&-V_{\text{transp}}^k u^k_\text{B,M}\\
	&=(k_u)_i^k(u^k_\text{B,M}-u^k_\text{T,M}),
\end{align*}
where $\chi_i^k$ was given by \eqref{chi_ik} and 
\begin{align*}
	(k_u)_i^k:=\frac{k_{u1}}{1+k_{u2} u_\text{T,M}^k+k_{u3}\displaystyle{\left(\int_{x_\text{M}}^{x_\text{N}} u^k \,dx\right)}^2}.
\end{align*}
Similarly, \eqref{sys-complete}\textsubscript{6} was discretised as 
\begin{align*}
	\frac{D_\varphi}{2\Delta x} \left(3 \varphi^k_\text{T,M} -4 \varphi^k_\text{T,M-1}+\varphi^k_\text{T,M-2}\right)&=k_\varphi(\varphi^k_\text{B,M}-\varphi^k_\text{T,M}),\\
	\frac{D_\varphi}{2\Delta x} \left(-3 \varphi^k_\text{B,M} +4 \varphi^k_\text{B,M+1}-\varphi^k_\text{B,M+2}\right)&=k_\varphi(\varphi^k_\text{B,M}-\varphi^k_\text{T,M}).
\end{align*}
In our numerical simulations we used $\Delta x=10^{-2}$ cm, while the interval $[x_\text{M}-\Delta x,x_\text{M}+\Delta x]$ centred on the membrane was discretised with $\Delta x_\text{f}=10^{-6}$ cm. Stability and non-negativity of numerical solutions were obtained by choosing $\Delta t=10^{-3}$ h.

%
%
%\begin{acknowledgements}
\bigskip
\bigskip
\subsection*{Acknowledgements}
We thank Gioconda Di Carluccio (IRCCS Istituto Nazionale Tumori ``Fondazione G. Pascale'', Naples, Italy) for her technical assistance. The assistance of the staff is gratefully appreciated. This work has been partially supported by the Italian Flagship Project InterOmics, by the PON01\_02460, and by AIRC (Associazione Italiana per la Ricerca sul Cancro) 2013, project 14225.
\subsection*{Author contributions}
EDC and RN conceived and designed the mathematical model. VI and MVC conceived and designed the experimental data. VI performed the experiments. EDC analysed the data, designed the numerical scheme, and performed the numerical simulations. CA, MFC contributed to analyse the data, to design the numerical scheme, and to perform the numerical simulations. EDC, MVC wrote the paper. VI, CA, MFC, RN contributed to the final version of the paper.
\bibliography{biblio_migration_model} % BibTeX database without .bib extension
\bibliographystyle{spbasic}

\newpage
\beginsupplement
\section*{Supplementary material}\label{sec:supplementary}

%\begin{figure}[htbp!]
%\centering
%\includegraphics[width=0.5\textwidth, angle=0]{figure/normalized_migration.pdf}
%%\caption{ (a) . (b) .}
%\label{fig:sperimentali2}   
%\end{figure}
%
%
%
%
%
%
%
%
%
%
%
%

\end{document}